\documentclass[hidelinks]{article}
\usepackage[total={7in,9in}]{geometry}
\usepackage{adjustbox} 
\usepackage{algorithm}
\usepackage{algpseudocode} 
\usepackage{float}
\usepackage{color}
\usepackage{dsfont} 
\usepackage{graphicx}
\usepackage{hyperref}
\usepackage{mathrsfs}
\usepackage{rotating}
\usepackage{wrapfig}
\usepackage{setspace}
\usepackage{xcolor} 
\usepackage{multirow}
\usepackage{booktabs,siunitx}
\usepackage{lipsum}
\usepackage{caption}
\usepackage{subcaption}
\usepackage{bigints}  
\usepackage{bm} 
\usepackage[numbers]{natbib}
\usepackage{authblk}
\providecommand{\keywords}[1]{\textbf{\textit{Keywords:}} #1}

\DeclareMathOperator*{\argmin}{arg\,min\,}
\DeclareMathOperator*{\argmax}{arg\,max\,}
\DeclareMathOperator*{\pr}{Pr}

\DeclareMathOperator{\ber}{Bernoulli}

\DeclareMathOperator{\expit}{expit}

\begin{document}

\title{Bayesian Machine Learning for Estimating Optimal Dynamic Treatment Regimes with Ordinal Outcomes}

\author[1]{Xinru Wang}

\author[2,3]{Tanujit Chakraborty}

\author[1,4,5$\ast$]{Bibhas Chakraborty}

\footnotesize{
\affil[1]{Centre for Quantitative Medicine, Duke-NUS Medical School, Singapore} 
\affil[2]{School of Data Science and Engineering, Sorbonne University Abu Dhabi, Abu Dhabi, United Arab Emirates} 
\affil[3]{Sorbonne Center For Artificial Intelligence, Sorbonne University, Paris, France} 
\affil[4]{Department of Statistics and Data Science, National University of Singapore, Singapore} 
\affil[5]{Department of Biostatistics and Bioinformatics, Duke University, Durham, USA} 
}

\date{}

\maketitle

\begin{abstract}
Dynamic treatment regimes (DTRs) are sequences of decision rules designed to tailor treatment based on patients' treatment history and evolving disease status. Ordinal outcomes frequently serve as primary endpoints in clinical trials and observational studies. However, constructing optimal DTRs for ordinal outcomes has been underexplored. This paper introduces a Bayesian machine learning (BML) framework to handle ordinal outcomes in the DTR setting. To deal with potential nonlinear associations between outcomes and predictors, we first introduce ordinal Bayesian additive regression trees (OBART), a new model that integrates the latent variable framework within the traditional Bayesian additive regression trees (BART). We then incorporate OBART into the BML to estimate optimal DTRs based on ordinal data and quantify the associated uncertainties. Extensive simulation studies are conducted to evaluate the performance of the proposed approach against existing methods. We demonstrate the application of the proposed approach using data from a smoking cessation trial and provide the OBART R package along with R code for implementation.  
\end{abstract}


\section{Introduction}
\label{sec:intro}

In real-life clinical practice dealing with chronic and relapsing diseases, such as mental illness \citep{nahum2015Health_psycho} and human immunodeficiency virus  (HIV) infection \citep{mustanski2020evaluation}, a single treatment may show heterogeneous effects for patients with different characteristics. To minimize unnecessary side effects and improve treatment efficacy, it is often preferable to offer treatments in multiple stages, adjusting dosage and intensity based on patients' evolving disease conditions. Dynamic treatment regimes (DTRs) are sequences of decision rules that formalize the multi-stage personalized treatment by taking the treatment histories and covariates as input and the recommended treatments as output \citep{murphy2005experimental}. 

In recent years, there are increasing methodological innovations for estimating optimal DTRs using sequential multiple assignment randomized trials (SMARTs) \citep{murphy2005experimental, wang2023sequential} or observational studies \citep{zhao2015new, wallace2015doubly, murray2018bayesian, tao2018tree}. For example, Wallace and Moodie (2015) \citep{wallace2015doubly} presented a dynamic weighted ordinary least squares regression (dWOLS) approach for estimating optimal DTRs for continuous data, offering both the doubly robust property of the G-estimation and the ease of implementation property of the Q-learning. Simoneau et al. (2020) \citep{simoneau2020estimating} extended the work of 
Wallace and Moodie (2015) \citep{wallace2015doubly} to survival endpoints subject to right-censoring. In addition to parametric or semiparametric models, Tao et al. (2018) \citep{tao2018tree} proposed more flexible and interpretable tree-based methods for estimating optimal DTRs. 

To the best of our knowledge, the majority of these methodologies focus on continuous, binary, or time-to-event outcomes. However, in many clinical trials, the primary endpoint is formalized as ordinal outcomes.
This is true for many SMARTs as well. One example is a two-stage smoking cessation trial \citep{strecher2008web} conducted by the Center for Health Communications Research (CHCR) at the University of Michigan. This smoking cessation trial, designed to help people quit smoking, consists of two sub-studies that together form a SMART. The first sub-study is the ProjectQuit (PQ) study conducted to find an optimal multi-component behavioral intervention to help adult smokers quit smoking and the second is the ForeverFree (FF) study conducted to help those who already quit stay quit and help those who failed at the initial stage with a second chance of receiving a beneficial treatment. Although the original studies used the binary quit status in the last seven days prior to the final survey as the primary outcome to evaluate the treatment, this outcome can be criticized as not measuring the long-term behavioral changes for an addiction like smoking; as such, many other secondary outcomes, e.g., the number of months not smoked by a subject over the study period, ranging from 0 to 6, were also collected \citep{chakraborty2009study}. To handle near-zero frequencies of some of the months in the distribution, a collapsed version of this variable, re-coded to three levels such as 0, 1, and 2, was used. Researchers may be interested in estimating the optimal decision rules based on such ordinal outcomes using six-month data instead of the binary quit status in the last seven days.

In such cases, the existing methods for constructing optimal DTRs can only be applied by transforming the ordinal data into continuous ones using scores for each category provided by relevant experts or dichotomizing the ordinal outcomes with more than two categories into two categories. Despite the simplicity of not requiring new statistical methodology, the potential drawback of the former approach is that the results may be sensitive to the specified scores, thus leading to biased estimates of the optimal DTRs, which will be more problematic when there is no consensus on the suitable numeric scores for each outcome category. Additionally, useful information may be wasted for the latter approach when dichotomizing the ordinal outcomes into binary outcomes. 

Various methodologies have been proposed to analyze ordinal data in non-DTR settings. Johnson and Albert (1999) \citep{Johnson1999} presented a comprehensive overview of regression models for ordinary data from both the frequentist and the Bayesian perspectives. However, limited literature focuses on ordinal data in the field of DTRs. Within the DTR literature, Ghosh et al. (2023) \citep{ghosh2023novel} proposed an easy-to-interpret measure of association, named generalized odds ratio (GOR), for comparing two embedded regimes in a SMART with ordinal outcomes. However, the GOR cannot be utilized when the research goal is to develop more tailored DTRs among a large set of potential DTRs. In addition, Jiang et al. (2024) \citep{jiang2024estimating} proposed a doubly robust dWOLS approach for constructing optimal DTRs for ordinal outcomes with household interference, i.e., one household member's treatment may affect another's outcome. However, as pointed out by Wallace and Moodie (2015) \citep{wallace2015doubly}, the standard bootstrap approach for variance estimation may be subject to non-regularity issues. 

Recently, Murray et al. (2018) \citep{murray2018bayesian} proposed a Bayesian machine learning (BML) approach, which innovatively combines Bayesian inferences with dynamic programming, to estimate an optimal DTR and quantify the uncertainties about the estimated parameters. Built upon their work, the current paper aims to propose a BML approach to make valid inferences about the optimal DTRs using data with ordinal outcomes. Our main contributions include presenting both frequentist and Bayesian perspectives for constructing optimal DTRs when dealing with ordinal outcomes, which are commonly used in medical settings but often overlooked. Specifically, we introduce an easy-to-implement Q-learning approach for ordinal data. To alleviate the non-regularity issues when quantifying the uncertainties of the parameters of interest in Q-learning \citep{chakraborty2013inference}, we propose to use the BML approach \citep{murray2018bayesian}, utilizing a data augmentation approach to account for all sources of uncertainties from stage-2 estimation when estimating optimal decision rules at stage 1. Built upon Chipman et al. (2010) \citep{chipman2010bart} and Tan and Roy (2019) \citep{tan2019}, we propose the Ordinal Bayesian Additive Regression Tree (OBART), and then incorporate it into the BML approach to capture nonlinear interactions between covariates and treatments and name it BML-OBART. We conduct extensive simulation studies to compare the candidate approaches mentioned above. We then apply the BML-OBART to a smoking cessation trial, offering a novel analysis of the SMART data with ordinal outcomes using Bayesian techniques, unlike previous studies focusing on analyzing binary outcomes using exclusively frequentist approaches \citep{chakraborty2010inference,zhao2015new}.  

The rest of the current paper is organized as follows. Section~\ref{sec:methods} provides the basic set-up, including data structures and notations in a multi-stage decision-making framework. Additionally, to better articulate our approach, we recapitulate the related work, i.e., Q-learning and BML approaches for continuous outcomes. In Section~\ref{sec:methodsOBML}, we introduce the proposed BML-OBART for estimating optimal DTRs with ordinal outcomes. We conduct simulation studies under various settings to evaluate the performance of the candidate approaches in Section~\ref{sec:simulation}. We then apply the proposed approaches to analyze data from a two-stage smoking cessation trial in Section~\ref{sec:application}. We conclude this paper with a comprehensive discussion in Section~\ref{sec:discussion}.

\section{Background and notations}
\label{sec:methods}

Without loss of generalizability, we focus on a two-stage decision-making problem with two available treatment options $A_j \in \mathcal{A}_j = \{ -1, 1\}$ at the $j$-th $(j = 1, 2)$ stage, where $A_j$ and $\mathcal{A}_j$ denote the received treatment and the treatment domain at stage $j$, respectively. Suppose there are $n$ independent and identically distributed participants' trajectories denoted by $(\bm{X}^{T}_{1}, A_{1}, \bm{X}^{T}_{2}, A_{2}, Y_2)$. The vector $\bm{X}_{j} \in \mathcal{X}_{j}$ consists of all the covariates collected before treatment $A_j$. Define $\bm{H}_{j} \in \mathcal{H}_{j}$ as the history data prior to the treatment at stage $j$, with $\bm{H}_1 = \bm{X}_1$, $\bm{H}_2 = (\bm{X}^{T}_1, A_1, \bm{X}^{T}_2)$, where $\mathcal{H}_{j}$ is the domain of history data prior to stage $j$. Let $Y_2$ be the observed outcome after stage 2. Note that in this paper, we only consider the cases without intermediate outcome $Y_1$. Hereafter, unless stated otherwise, we use uppercase letters to represent random variables, lowercase letters to represent the corresponding observed values, and bold letters to represent vectors when no confusion exists. 

We adopt the potential outcomes framework proposed by Rubin (1974) \citep{rubin1974estimating}. Let $Y_2(a_1, a_2)$ denote the potential outcome that would be observed when $A_1 = a_1$ and $A_2 = a_2$, and let $\bm{X}_2(a_1)$ represent the potential tailoring variables at stage 2 when $A_1 = a_1$. We assume the following: 1) Consistency, which asserts that the observed variables correspond to the potential variables under the treatment received. Consequently, if we observe $(\bm{x}_1^T, a_1, \bm{x}_2^T, a_2, y_2)$, then $\bm{x}_2^T = \bm{x}_2^T(a_1)$ and $y_2 = y_2(a_1, a_2)$; 2) Exchangeability, which states that $\pr(A_j = a_j | \bm{H}_2)$ is independent of the potential outcomes; and 3) Positivity, which ensures that $\pr(A_j = a_j | \bm{H}_2) > 0$. These assumptions enable the estimation of causal effects at each stage based on the observed data. For a better exposition of our proposed approach, in this section, we first introduce Q-learning and BML for continuous outcomes, i.e., $Y_2 \in \mathcal{R}$, where higher values correspond to better outcomes. We define the two-stage DTRs as $\bm{d} = (d_1, d_2) \in \mathcal{D}$, where $d_j \in \mathcal{D}_j$ is a function that maps $\mathcal{H}_j$, i.e., the domain of the history data, to $\mathcal{A}_j$, i.e., the domain of treatment at stage $j$, and $\mathcal{D}_j$ is the domain of all possible decision rules at stage $j$. Our goal is to find the optimal DTR $\bm{d}^{\ast}$ that maximizes the expected outcome. 

\subsection{Estimating optimal DTRs with continuous outcomes by Q-learning}
\label{sec:methods1}

For continuous outcomes, Q-learning is a commonly used and easy-to-implement method for estimating optimal DTRs. In summary, the optimal stage-2 decision rule $d_2^{\ast}$ and the corresponding potential outcome $y_2(a_1, d_2^{\ast})$ under the optimal decision rule are identified by building a stage-2 model for $y_2(a_1, a_2)$ based on the observed data with $y_2$ as response variable and $\bm{H}_2$ and $A_2$ as predictors. The potential outcome $y_2(a_1, d_2^{\ast})$ is then used as the response variable to estimate optimal decision rule at stage 1. For observations whose observed stage-2 treatments match the optimal decision rules, the potential outcomes $y_2(a_1, d_2^{\ast})$ are observed and equal to $y_2$. However, for those who do not receive the optimal decision rule, the potential outcome needs to be imputed by the stage-2 model. Specifically, we define the optimal decision rules at each stage as the one that maximizes the expected outcome \citep{moodie2012q}, i.e.,  
\begin{equation}
\label{eq_qlearning1}
d_2^{\ast}(\bm{h}_2) = \underset{a_2 \in \mathcal{A}_2}{\argmax} Q_2(\bm{h}_2, a_2) = \underset{a_2 \in \mathcal{A}_2}{\argmax}  E[Y_2|\bm{h}_2, a_2], \quad \forall \; \bm{h}_2 \in \mathcal{H}_2,
\end{equation}
\begin{equation}
\label{eq_qlearning2}
d_1^{\ast}(\bm{h}_1) = \underset{a_1 \in \mathcal{A}_1}{\argmax} Q_1(\bm{h}_1, a_1) =  \underset{a_1 \in \mathcal{A}_1}{\argmax}  E[Y_1|\bm{h}_1,a_1], \quad \forall \; \bm{h}_1 \in \mathcal{H}_1,
\end{equation}
where $Q_j(\bm{h}_j, a_j) = E[Y_j|\bm{h}_j, a_j]$ is defined as the Q-function for stage $j$; the pseudo-outcome for stage 1 is defined as $Y_1=\underset{a_2 \in \mathcal{A}_2}{\max} E[Y_2|\bm{h}_2, a_2] = Y_2(a_1, d_2^{\ast})$, which is the expected outcome under the optimal treatment at stage 2.  When the Q-functions are specified as linear models, i.e., $Q_j(\bm{h}_j, a_j|\bm{\beta}_j,\bm{\zeta}_j) = \bm{\beta}_j^{T} \bm{h}_{j0}+\bm{\zeta}_j^T \bm{h}_{j1} a_j $, where $\bm{h}_{j0}$ and $\bm{h}_{j1}$ are two potentially different features of the history data $\bm{h}_j$, the vectors of parameters $\bm{\beta}_j$ and $\bm{\zeta}_j$ can be estimated by the ordinary least squares (OLS). 
The estimated optimal decision rule at stage $j$ is thus $\hat{d}^{\ast}_j(\bm{h}_j)=\underset{a_j \in \mathcal{A}_j}{\argmax} Q_j(\bm{h}_j,a_j|\hat{\bm{\beta}}_j, \hat{\bm{\zeta}}_j)$ and the estimated optimal DTR is $\hat{\bm{d}}^{\ast}=(\hat{d}_1^{\ast}, \hat{d}_2^{\ast})$. Although Q-learning is interpretable and easy to implement, it is sensitive to model mis-specifications and subject to non-regularity issues. \citep{chakraborty2013inference}. 

\subsection{Estimating optimal DTRs for continuous outcomes by Bayesian Machine Learning (BML)}
\label{sec:methods2}

\sloppy The BML approach proposed by Murray et al. (2018) \citep{murray2018bayesian} is a novel way that combines Bayesian inference with dynamic programming and allows for the quantification of uncertainties in estimated parameters. The overall procedure for the BML approach involves a series of Bayesian models in a backward and recursive manner to construct stage-specific optimal decision rules. Under the consistency assumptions, we directly use $Y_2$ and $\bm{X}_2$ instead of $Y_2(a_1,a_2)$ and $\bm{X}_2(a_1)$ in the following sections. For stage 2, the distribution of the outcome conditional on the historical characteristics and treatments can be defined as 
\begin{equation*}
\label{eq_bmlc1}
\begin{aligned}
    Y_2 | \bm{H}_2, A_2, \bm{\beta}_2,\bm{\zeta}_2 &\sim
q_2(Y_2|\bm{H}_2,A_2,\bm{\beta}_2,\bm{\zeta}_2), \quad
    \bm{\beta}_2,\bm{\zeta}_2 & \sim p_2^{(0)}(\bm{\beta}_2,\bm{\zeta}_2),
\end{aligned} 
\end{equation*}
where $p_2^{(0)}(\bm{\beta}_2,\bm{\zeta}_2)$ is the prior distribution of the unknown model parameters $\bm{\beta}_2,\bm{\zeta}_2$. The posterior distribution for $\bm{\beta}_2,\bm{\zeta}_2$ is thus 
\begin{equation*}
\label{eq_bmlc2}
p_{2,n}(\bm{\beta}_2,\bm{\zeta}_2|\bm{O}_n) \propto  p_2^{(0)}(\bm{\beta}_2,\bm{\zeta}_2) \times \prod_{i=1}^n q_2(Y_{2i}|\bm{H}_{2i},A_{2i},\bm{\beta}_2,\bm{\zeta}_2),
\end{equation*}
where $\bm{O}_n = \{\bm{X}^T_{1i}, A_{1i}, \bm{X}^T_{2i}, A_{2i}, Y_{2i}\}^n_{i=1}$ denotes the observed data for $n$ participants. The optimal decision rule at stage 2 is then defined as 
\begin{equation}
\label{eq_bmlc3}
d_2^{\ast}(\bm{h}_2) = \underset{a_2 \in \mathcal{A}_2}{\argmax } E[Y_2|\bm{H}_2,a_2,\bm{O}_n], \quad \forall \; \bm{H}_2 \in \mathcal{H}_2,
\end{equation}
where the expectation is taken with respect to the predictive distribution of $Y_2$, i.e., $q_{2,n}(Y_2|\bm{H}_2,A_2,\bm{O}_n) = \int q_2(Y_2|\bm{H}_2,A_2,\bm{\beta}_2,\bm{\zeta}_2) p_{2,n}(\bm{\beta}_2,\bm{\zeta}_2|\bm{O}_n) d (\bm{\beta}_2,\bm{\zeta}_2).$
As noted by Murray et al. (2018) \citep{murray2018bayesian}, the inference procedure at stage 2 can be implemented using standard Bayesian models as it only requires the observed outcomes $Y_2$ as response variables and all the covariates $\bm{H}_2$ and $A_2$ as predictors in the model. However, for stage 1, the stage-1 pseudo-outcome assuming the optimal stage-2 treatment is unknown. Furthermore, the optimal stage-2 treatment itself is unknown and is estimated based on the observed data. Consequently, additional efforts are required to account for the uncertainties in estimating the optimal decision rules at stage 2 and the potentially unknown outcome under the estimated optimal decision rules. The posterior distribution of the stage-1 model parameters, accounting for these two sources of uncertainties, is expressed as 
\begin{equation}
\label{eq_bmlc4}
\resizebox{0.8\textwidth}{!}{
$\begin{aligned}
    p_{1,n}(\bm{\beta}_1,\bm{\zeta}_1|\bm{O}_n) & \propto p_1^{(0)}(\bm{\beta}_1,\bm{\zeta}_1) \times 
    \bigint \left\{ \prod_{i: A_{2i}=\hat{d}^{\ast}_2(\bm{H}_{2i}|\bm{\beta}_2,\bm{\zeta}_2)} q_1(Y_{2i}|\bm{H}_{1i},A_{1i},\bm{\beta}_1,\bm{\zeta}_1) \right.\\
    & \left. \times \int \prod_{i: A_{2i} \neq \hat{d}^{\ast}_2(\bm{H}_{2i}|\bm{\beta}_2,\bm{\zeta}_2)} q_1(Y_{1i}^{mis}|\bm{H}_{1i},A_{1i},\bm{\beta}_1,\bm{\zeta}_1) q_{2,n}(Y_{1i}^{mis}|\bm{H}_{2i},A_{2i},\bm{O}_n) d Y_{1i}^{mis} \right\} \times p_{2,n}(\bm{\beta}_2,\bm{\zeta}_2|\bm{O}_n) d(\bm{\beta}_2,\bm{\zeta}_2),
\end{aligned}$
}
\end{equation}
where $p_1^{(0)}(\bm{\beta}_1,\bm{\zeta}_1)$ is the prior distribution of the parameters in the stage-1 regression model; $Y_{1i}^{mis}$ is the stage-1 pseudo outcome $Y_2(a_1, d_2^{\ast})$, which requires imputation for those who do not receive the estimated optimal stage-2 decision rule; and $\hat{d}_2^{\ast}(\bm{H}_2|\bm{\beta}_2,\bm{\zeta}_2)$ is the optimal stage-2 decision rule under parameters $\bm{\beta}_2,\bm{\zeta}_2$. For those who receive the estimated optimal stage-2 decision rule, i.e., $\{i: A_{2i} = \hat{d}^{\ast}_2(\bm{H}_{2i}|\bm{\beta}_2,\bm{\zeta}_2)\}$, we set $Y_{1i}^{mis} = Y_{2i}$. In Eq.(\ref{eq_bmlc4}), $Y_{1i}^{mis}$ appears in two contexts: as part of the likelihood function $q_1(\cdot)$ for the response variable and within $q_{2,n}(\cdot)$, which is treated as a posterior distribution to integrate over $Y_{1i}^{mis}$. 

The primary difference between the posterior distribution of the stage-1 and stage-2 parameters is that the stage-1 posterior distribution should account for the uncertainties in the optimal decision rules at stage 2 and the corresponding potential outcome. Recall that the stage-1 pseudo outcome $Y_1$ is defined as the potential outcome under the optimal decision rule at stage 2 $d_2^{\ast}(\bm{H}_2)$ and $Y_1$ is then used to estimate the optimal decision rule at stage 1. If the true value of $Y_1$ were known, conventional Bayesian models could be used to estimate stage-1 parameters. However, in reality, $Y_1$ is unknown due to two main sources of uncertainties:
1) the optimal decision rule $d_2^{\ast}(\bm{H}_2)$ is unknown and needs to be estimated based on stage-2 data; and 2) even with $d_2^{\ast}(\bm{H}_2)$ known, the true value of $Y_1$ under $d_2^{\ast}(\bm{H}_2)$ remains unobserved for individuals who do not receive $d_2^{\ast}(\bm{H}_2)$. These uncertainties need to be taken into account when estimating the posterior distributions of stage-1 parameters. Since the posterior distribution of $d_2^{\ast}(\bm{H}_2)$ depends on the posterior distribution of $(\bm{\beta}_2,\bm{\zeta}_2)$, we integrate the posterior distribution of $(\bm{\beta}_2,\bm{\zeta}_2)$ in Eq.(\ref{eq_bmlc4}) to account for the uncertainties of stage-2 optimal decision rules. We then integrate the predictive distribution of the outcome $Y_1^{mis}$ for those who do not receive the corresponding stage-2 optimal decision rule to take into account the uncertainties in estimating $Y_1$. The optimal decision rule at stage 1 is thus 
\begin{equation}
\label{eq_bmlc5}
d_1^{\ast}(\bm{h}_1) = \underset{a_1 \in \mathcal{A}_1}{\argmax } E[Y_1|\bm{H}_1,a_1,\bm{O}_n], \quad \forall \; \bm{H}_1 \in \mathcal{H}_1,
\end{equation}
where the expectation is taken with respect to the predictive distribution of $Y_1$, i.e., $$q_{1,n}(Y_1|\bm{H}_1,A_1,\bm{O}_n) = \int q_1(Y_1|\bm{H}_1,A_1,\bm{\beta}_1,\bm{\zeta}_1) p_{1,n}(\bm{\beta}_1,\bm{\zeta}_1,\bm{O}_n) d (\bm{\beta}_1,\bm{\zeta}_1).$$ To simplify the sampling procedure for deriving the posterior distribution for the stage-1 parameters, Murray et al. (2018) \citep{murray2018bayesian} proposed the \textit{backward induction Gibbs (BIG)} sampler to draw posterior samples of model parameters, which significantly simplifies the computation through Bayesian data augmentation and makes the sampling from the stage-1 posterior distribution practically feasible. We will omit the detailed steps for the BIG sampler here and introduce the BIG sampler for ordinal outcomes in the next section.

\section{Extending Q-learning and BML for ordinal outcomes}
\label{sec:methodsOBML}

Apart from continuous and binary outcomes, ordinal outcomes are commonly used in clinical settings to assess treatment efficacy. In this section, we aim to extend Q-learning and BML to ordinal outcome $Y_2$ with $K$ levels. Without loss of generality, we assume that higher levels are better but the quantitative differences between any two levels are unknown. 

We assume a latent normal distributed variable at stage $j$: $Z_j = f_j(\bm{H}_j, A_j) + \epsilon_j, \quad \epsilon_j \sim N(0,1)$ \citep{Johnson1999}. For parametric models, $E[Z_j|\bm{h}_j, a_j] = f_j(\bm{h}_j, a_j)$ can be defined as $f_j(\bm{h}_{j}, a_{j}) = \bm{\beta}_j^T \bm{h}_{j1} + \bm{\zeta}_j^T a_j \bm{h}_{j2}$. The latent continuous outcome is categorized into ordinal outcomes based on cutoff values $\bm{\gamma}_j = (\gamma_{j,0}, \dots, \gamma_{j,K} )$ at stage $j$, with $\gamma_{j,0}=-\infty, \gamma_{j,K}=\infty$. If $\gamma_{j,k-1} < Z_j \leq \gamma_{j,k}$, then $Y_j = k$. The cumulative probability of being less than or equal to category $k$ at stage $j$ given the history $\bm{H}_j$ and treatment $A_2$ is defined as $\theta_{j,k}(\bm{H}_j, A_2)=\pr(Y_j \leq k|\bm{H}_2, A_2) = \Phi(\gamma_{j,k} - f_j(\bm{H}_2, A_2))$, and the probability of being category $k$ is denoted as $\pi_{j,k}(\bm{H}_2, A_2)= \theta_{j,k}(\bm{H}_2, A_2) - \theta_{j,k-1}(\bm{H}_2, A_2)$, where $\Phi$ is the cumulative distribution function (CDF) of the standard normal distribution.

\subsection{Q-learning for estimating optimal DTRs with ordinal outcomes}
\label{sec:methodsOBML1}

For continuous outcomes, the optimal decision rule at stage $j$ is the one that maximizes the expected outcome. However, for ordinal outcomes, we define the optimal decision rule at stage $j$ as the one maximizing the latent continuous outcome $Z_j$ at stage $j$, i.e., 
\begin{equation}
\label{eq_qlearningo1}
d_j^{\ast}(\bm{h}_j) = \underset{a_j \in \mathcal{A}_j}{\argmax} \  Q_j(\bm{h}_j, a_j) =\underset{a_j \in \mathcal{A}_j}{\argmax} E[Z_j|\bm{h}_j, a_j] .
\end{equation}
By maximizing the latent continuous outcome $Z_j$, the probability of falling into higher catgeories for outcome $Y_2$ is maximized. When a linear model is specified for the latent continuous variable $Z_j$, 
\begin{equation}
\label{eq_qlearningo2}
E[Z_j|\bm{h}_j, a_j] = f_j(\bm{h}_j, a_j) = \bm{\beta}_j^T \bm{h}_{j1} + \bm{\zeta}_j^T a_j  \bm{h}_{j2},
\end{equation}
the optimal treatment $d_j^{\ast}(\bm{h}_j) = 1$ if $\psi_{j}(\bm{h}_j) = f_j(\bm{h}_j, 1) - f_j(\bm{h}_j, -1) = 2\bm{\zeta}_j^T \bm{h}_{j2} > 0$; otherwise, $d_2^{\ast}(\bm{h}_{j}) = -1$. Recall that for continuous outcomes, the pseudo outcome at stage 1 is defined as the expected outcome under the optimal decision rule at stage 2. However, for ordinal outcomes, we define the ordinal pseudo outcome $Y_1$ as a random variable following a multinomial distribution $q_2(Y_2|\bm{h}_2, d_2^{\ast}(\bm{h}_{2}))$ under the optimal decision rule $d_2^{\ast}(\bm{h}_{2})$ at stage 2, with probability for level $k$ at stage $j$ denoted as $\pi_{2,k}(\bm{h}_{2}, d_2^{\ast}(\bm{h}_{2}))$. We first build models, e.g., ordered probit models \citep{albert1993bayesian}, to estimate the cut-off values $\bm{\gamma}_j$ and $\bm{Z}_j$ at each stage and then derive the optimal decision rules based on the estimated model parameters $(\hat{\bm{\beta}}_j, \hat{\bm{\zeta}}_j,\hat{\bm{\gamma}}_j)$. The detailed steps using Q-learning are provided in Algorithm~\ref{algorithm_qlearning}. 

\begin{algorithm}
\caption{Estimating Optimal DTRs with Ordinal Data using Q-learning (frequentist method).}
\label{algorithm_qlearning}
\begin{algorithmic}[1]

\Require 
Observed data $\bm{O}_n = {(\bm{X}^T_{1i}, A_{1i}, \bm{X}^T_{2i}, A_{2i}, Y_{2i})}_{i=1}^n$, the number of repetitions $R^{ql}$.
\Statex \textbf{Steps:}
   \State \sloppy  Estimate $\hat{\bm{\beta}}_2, \hat{\bm{\zeta}}_2, \hat{\bm{\gamma}}_2$ based on observed data $\bm{O}_n = \{(\bm{X}_{1i}, A_{1i}, \bm{X}_{2i}, A_{2i}, Y_{2i})\}_{i=1}^n$. The estimated optimal decision rule is $\hat{d}_2^{\ast}(\bm{H}_{2i}) = \underset{a_2 \in \mathcal{A}_2}{\arg\max} \hat{f}_2(\bm{H}_{2i}, A_{2i})$.
    \State Generate the candidate ordinal pseudo outcome $\tilde{Y}_{1i}$ based on a multinomial distribution with probability for level $k$ denoted as $\hat{\pi}_{2,k}(\bm{H}_{2i}, \hat{d}_2^{\ast}(\bm{H}_{2i}))$. The pseudo outcome at stage 1 is $\hat{Y}_{1i} = I(A_{2i} = \hat{d}_2^{\ast}(\bm{H}_{2i}))Y_{2i} + (1-I(A_{2i} = \hat{d}_2^{\ast}(\bm{H}_{2i}))) \tilde{Y}_{1i}$, where $I(\cdot)$ is the indicator function.
    \State Estimate $\bm{\beta}_1, \bm{\zeta}_1$ based on $\{(\bm{X}_{1i}, A_{1i}, \hat{Y}_{1i})\}_{i=1}^n$.
    \State Repeat Step 2 and Step 3 $R^{ql}$ times to get multiple values of $\hat{\bm{\beta}}_1^{(r)}, \hat{\bm{\zeta}}_1^{(r)}$, $r=1, \dots, R^{ql}$. The final estimates are $\hat{\bm{\beta}}_1 = \sum_{r=1}^{R^{ql}} \hat{\bm{\beta}}_1^{(r)}/R^{ql}$ and $\hat{\bm{\zeta}}_1 = \sum_{r=1}^{R^{ql}} \hat{\bm{\zeta}}_1^{(r)}/R^{ql}$. The estimated optimal decision rule at stage 1 is $\hat{d}_1^{\ast}(\bm{H}_{1i}) = \underset{a_1 \in \mathcal{A}_1}{\arg\max} \hat{f}_1(\bm{H}_{1i}, A_{1i})$.
\Ensure Estimated optimal DTR $ \hat{d}^{\ast} =(\hat{d}_1^{\ast},\hat{d}_2^{\ast})$, estimated model parameter $\hat{\bm{\beta}}_1, \hat{\bm{\zeta}}_1, \hat{\bm{\beta}}_2, \hat{\bm{\zeta}}_2, \hat{\bm{\gamma}}_2$.
\end{algorithmic}
\end{algorithm}

To reduce randomness in generating stage-1 pseudo-outcomes and enhance efficiency, we generate $R^{ql}$ samples based on the estimated multinomial distribution under the optimal decision rule \citep{jiang2024estimating}. The estimation procedure is then applied to these $R^{ql}$ samples to obtain the average values of the model parameters. A sufficient number should be used to ensure relatively stable estimates. Similar to Q-learning for continuous outcomes, the Q-learning procedure for ordinal outcomes lacks robustness to model mis-specifications. In addition, quantifying the uncertainties in estimating stage-1 parameters can be difficult, especially when there are no significant treatment effects at stage 2 \citep{chakraborty2013inference}. 

\subsection{The BML approach for estimating optimal DTRs with ordinal data}
\label{sec:methodsOBML2}

In this subsection, we extend the BML approach for ordinal outcomes. For stage 2, the standard Bayesian ordered probit model (BP) can be implemented directly using the `stan\_polr' function within the \textit{rstanarm} package \citep{rstan2024}. For stage 1, BIG sampler is employed to draw posterior samples. The detailed algorithm incorporating the Bayesian ordered probit model into the BML framework is summarized in Algorithm~\ref{algorithm_bml-bp}.

\begin{algorithm}
\caption{Estimating Optimal DTRs with Ordinal Data using BML-BP}
\label{algorithm_bml-bp}
\begin{algorithmic}[1]
\Require 
Observed data $\bm{O}_n = {(\bm{X}^T_{1i}, A_{1i}, \bm{X}^T_{2i}, A_{2i}, Y_{2i})}_{i=1}^n$, the number of posterior draws $R^{bml}$. 

\Statex \textbf{Steps:}
    \State Obtain the posterior distribution $p_{2,n}(\bm{\beta}_2,\bm{\zeta}_2,\bm{\gamma}_2|\bm{O}_n)$ based on observed data. The optimal decision rule is selected based on Eq.~(\ref{eq_bmlc3}).
    
    \State Draw $R^{bml}$ samples $(\bm{\beta}_2,\bm{\zeta}_2,\bm{\gamma}_2)^{(r)}, r=1, \dots, R^{bml}$ from $p_{2,n}(\bm{\beta}_2,\bm{\zeta}_2,\bm{\gamma}_2|\bm{O}_n)$. For each sample $r$, get the optimal decision rule $\hat{d}_2^{\ast(r)}(\bm{H}_{2i}|\bm{\beta}_2^{(r)}, \bm{\zeta}_2^{(r)})$.

    \State For each $i$ in sample $r$, generate $\tilde{Y}_{1i}^{(r)}$ by drawing from $q_{2,n}(Y_{2i}|\bm{H}_{2i},\hat{d}^{\ast(r)}_2(\bm{H}_{2i}| \bm{\beta}_2^{(r)},\bm{\zeta}_2^{(r)}), \bm{O}_n)$. Set $\hat{Y}_{1i}^{(r)} = Y_{2i}$ if $A_{2i}=\hat{d}_2^{\ast(r)}(\bm{H}_{2i}| \bm{\beta}_2^{(r)},\bm{\zeta}_2^{(r)})$, else set $\hat{Y}_{1i}^{(r)}=\tilde{Y}_{1i}^{(r)}$. 
    \State For each imputed dataset $r$, ontain the posterior distribution of stage-1 model parameters $p^{(r)}_{1,n}(\bm{\beta}_1,\bm{\zeta}_1,\bm{\gamma}_1|\bm{O}_n)$, from which we draw one sample $(\bm{\beta}^{(r)}_1,\bm{\zeta}^{(r)}_1,\bm{\gamma}^{(r)}_1)$. Combine all $R^{bml}$ draws to represent the final posterior distribution of $(\bm{\beta}_1,\bm{\zeta}_1,\bm{\gamma}_1)$.

    \State Get the optimal decision rules at stage 1 $\hat{d}^{\ast}_1(\bm{H}_{1i})$ based on Eq.~(\ref{eq_qlearningo1})

    \Ensure Estimated optimal DTR $(\hat{d}_1^{\ast},\hat{d}_2^{\ast})$, estimated model parameters $\hat{\bm{\beta}}_1, \hat{\bm{\zeta}}_1, \hat{\bm{\beta}}_2, \hat{\bm{\zeta}}_2, \hat{\bm{\gamma}}_2$.
\end{algorithmic}
\end{algorithm}

In brief, step 1 in Algorithm~\ref{algorithm_bml-bp} focuses on finding the optimal decision rules at stage 2 by obtaining the posterior distribution of stage-2 model parameters using Bayesian ordered probit (BP) models, while steps 2-5 focus on estimating the optimal decision rules at stage 1. Specifically, steps 2 and 3 draw multiple samples from the posterior distribution of stage-2 model parameters. Each sample corresponds to an optimal decision rule at stage 2 and the corresponding stage-1 pseudo-outcome. This iterative procedure generates multiple imputed stage-1 datasets. In step 4, for each imputed stage-1 dataset, we build a BP model to estimate the posterior distribution of stage-1 parameters and draw a single sample. The collection of samples based on these imputed stage-1 datasets represents the final posterior distribution of the stage-1 model parameters, which are then used to estimate the optimal decision rule at stage 1. 

Unlike Q-learning, the BML approach can quantify the uncertainty of the estimated parameters using the posterior distributions without incurring additional costs. However, similar to Q-learning, parametric Bayesian models may be sensitive to model mis-specifications, especially when dealing with complex and nonlinear data. To alleviate this problem, we consider integrating the non-parametric BART model into the BML framework. BART is a Bayesian sum-of-trees model developed by  Chipman et al. (2010) \citep{chipman2010bart} that can handle nonlinear main effects and multiple-way interactions between predictors. For continuous outcomes $Y$ and $p$-dimensional covariates $\bm{x} = (x_1,\dots,x_p)$, we define
\begin{equation}
\begin{aligned}
    Y &= E[Y|x_1,\dots,x_p] + \epsilon = f(\bm{x}) + \epsilon, \quad \epsilon \sim N(0,\sigma^2).
\end{aligned} 
\end{equation}

The aim of BART is to estimate $f(\bm{x})$, which is a sum of $M$ regression trees $f(\bm{x}) \equiv \sum_{m=1}^M f(\bm{x}|\mathcal{T}_m,\mathcal{M}_m)$, where $\mathcal{T}_m$ denotes the $m$-th binary tree, $\mathcal{M}_m = (\mu_{m1}, \dots, \mu_{m b_m})$ denotes the set of values of terminal nodes associated with $\mathcal{T}_m$, and $b_m$ is the number of terminal nodes for the $m$-th binary tree. Chipman et al. (2010) \citep{chipman2010bart} simplified the prior specification assuming that the components of each tree are independent and unrelated to the error term $\epsilon$. Under this assumption, $p[(\mathcal{T}_1,\mathcal{M}_1),\dots,(\mathcal{T}_M,\mathcal{M}_M),\sigma] =\left[ \prod_{m=1}^M \left\{ \prod_{i=1}^{b_m} p(\mu_{mi} | \mathcal{T}_m) \right\} p(\mathcal{T}_m) \right] p(\sigma)$. The priors for $\mu_{mi} | \mathcal{T}_m$ and $p(\sigma)$ are usually specified as $\mu_{mi} | \mathcal{T}_m \sim N(\mu_{\mu}, \sigma_{\mu}^2)$ and $\sigma^2 \sim IG(\frac{\nu}{2}, \frac{\nu \lambda}{2})$, respectively, where $IG(a_{\sigma},b_{\sigma})$ is the Inverse-Gamma distribution with shape parameter $a_{\sigma}$ and rate parameter $b_{\sigma}$. The prior for $p(T_m)$ involves: i) the probability of a node at depth $\rho = 0,1,2,\dots$ being non-terminal, specified by $\alpha(1+\rho)^{-\beta}$, where $\alpha \in (0,1)$ and $\beta \in [0,\infty)$; ii) the probability of being selected as a splitting variable; iii) for the selected splitting variable, the probability of splitting rules. The default values of the hyperparameters are provided by Tan and Roy (2019) \citep{tan2019}. Specifically, for continuous outcomes, the number of trees $M=200$, $b=2$, $\alpha = 0.95$, $\beta = 2$, and $\nu = 3$. The two-step Gibbs sampling \citep{albert1993bayesian} is employed to draw samples from the full conditional posterior distributions $p(\{\mathcal{T}_m, \mathcal{M}_m\}_{m=1}^M | \sigma, y)$ and $p(\sigma | \{\mathcal{T}_m, \mathcal{M}_m\}_{m=1}^M, y)$. Details about the MCMC sampling algorithms for BART can be found in Chipman et al. (2010) \citep{chipman2010bart} and Tan and Roy (2019) \citep{tan2019}.

In this paper, we propose ordinal BART to deal with ordinal outcomes, and call it OBART. Specifically, to avoid over-parameterization, the first cut-off parameter $\gamma_1$ is fixed to be $0$. The prior distribution for the tree structure is specified similarly to traditional BART introduced before \citep{chipman2010bart, tan2019}. The hyperparameter for the prior distribution of $\mu_{mi}$ is $\sigma_{\mu} = \frac{3}{b \sqrt{M}}$ and uniform noninformative priors are used for the cutoff parameters $\bm{\gamma}_j$. The posterior draws of trees $\{(\mathcal{T}_m, \mathcal{M}_m)\}_{m=1}^M$ can be obtained similarly to the traditional BART with the latent continuous variable $Z_{ji}$ as outcome variables and $\bm{H}_{ji}$ and $A_{ji}$ as predictors. A simple Gibbs sampling approach to draw posterior distributions of model parameters may be possible using conditional posterior distributions. Specifically, the cut-off value $\bm{\gamma}_{j,k}$ follows a truncated uniform distribution with interval $(\max_{Y_{ji}=k} Z_{ji}, \min_{Y_{ji}=k+1} Z_{ji})$. Furthermore, given $\bm{\gamma}_j$, $\bm{H}_{ji}$, $A_{ji}$ and $\{(\mathcal{T}, \mathcal{M})\}_{m=1}^M$, the latent variable $Z_{ji}$ follows a truncated normal distribution with unit variance. 

However, as pointed out by Albert and Chib (1993) \citep{albert1993bayesian}, in cases with many observations in adjacent categories, the resulting interval for $\bm{\gamma}_j$ 
from the Gibbs sampling procedure tends to be very narrow, leading to minimal movements and a slow convergence rate. Therefore, we employ a hybrid Metropolis-Hastings/Gibbs sampler \citep{cowles1996accelerating}. The sampling procedure for $\{(\mathcal{T}_m, \mathcal{M}_m)\}_{m=1}^M$ is the same as the traditional BART \citep{chipman2010bart}. However, the conditional posterior distribution of $\{\bm{Z}_{j}, \bm{\gamma}_j\}$ is factored into $p_{j,n}(\bm{Z}_{j} | \bm{Y}_j, \bm{\gamma}_j, \{(\mathcal{T}_m, \mathcal{M}_m)\}_{m=1}^M)$ and $p_{j,n}(\bm{\gamma}_j|\bm{Y}_j,\{(\mathcal{T}_m, \mathcal{M}_m)\}_{m=1}^M)$, where $p_{j,n}(\bm{\gamma}_j|\bm{Y}_j,\{(\mathcal{T}_m, \mathcal{M}_m)\}_{m=1}^M) \propto \prod_{i=1}^n \Phi(\gamma_{j,Y_{ji}} - f(\bm{H}_{ji}, A_{ji})) - \Phi(\gamma_{j,Y_{ji}-1} - f(\bm{H}_{ji}, A_{ji}))$.

Algorithm~\ref{algorithm_obart} provides the detailed steps to draw samples from the conditional posterior distributions. Note that $\sigma_{MH} = 0.5/K$ is set such that the acceptance ratio is within the desired range $(0.25-0.5)$, where $K$ is the number of categories. Adjustments should be made if the acceptance rate is below or above the desired range. Algorithm~\ref{algorithm_obart} is then incorporated into the step 1 and step 4 in Algorithm~\ref{algorithm_bml-bp} to estimate the optimal decision rules. We provide the package \textit{OBART} based on the existing \textit{BART} package \citep{bart2024} and the example R code for BML-OBART for estimating optimal DTRs with ordinal data at \url{https://github.com/cherylwaal/OBART} and \url{https://github.com/cherylwaal/BMLordinal}, respectively. 

\begin{algorithm}
\caption{Algorithm for OBART using a hybrid Metropolis-Hastings/Gibbs sampler}
\label{algorithm_obart}
\begin{algorithmic}[1]
\Require The initial value $Z_{ji}^{(0)}$ and $\bm{\gamma}_j^{(0)}$, $\sigma_{MH} = 0.5/K$, observed data $\bm{O}_n = (\bm{X}^T_{1i}, A_{1i}, \bm{X}^T_{2i}, A_{2i}, Y_{2i})_{i=1}^n$, the posterior draws $R^{ob}$.
\Statex \textbf{Steps:}
    \State Initialize $Z_{ji}^{(0)}$, $\bm{\gamma}_j^{(0)}$; set $t=1$. 
    \State Sample $M$ trees $(\mathcal{T}^{(t)}_m, \mathcal{M}^{(t)}_m), m=1, \dots, M$ using $Z^{(t-1)}_{ji}$ as response variables and $\bm{H}_{ji}, A_{ji}$ as predictors following the same steps as those in BART for continuous outcomes.
    \State For $k=2,\dots,K$, sample $\dot{\gamma}_{j,k} \sim N(\gamma_{j,k}^{(t-1)}, \sigma^2_{MH})I(\dot{\gamma}_{j,k-1}, \gamma_{j,k+1}^{(t-1)})$, compute the acceptance ratio (AR)
            \begin{equation}
            \scalebox{0.7}{
           $
                \begin{aligned}
                    AR&= \prod_{i=1}^n \frac{\Phi(\dot{\gamma}_{j,Y_{ji}} - f_{ji}^{(t)}) - \Phi(\dot{\gamma}_{j,Y_{ji}-1} - f_{ji}^{(t)})}{\Phi(\gamma_{j,Y_{ji}}^{(t-1)} - f_{ji}^{(t)}) - \Phi(\gamma_{j,Y_{ji}-1}^{(t-1)} - f_{ji}^{(t)})} \\
                    & \times \prod_{k=2}^{K-1} \frac{\Phi((\gamma_{j,k+1}^{(t-1)} - \gamma_{j,k}^{(t-1)})/\sigma_{MH}) - \Phi((\dot{\gamma}_{j,k-1} - \gamma_{j,k}^{(t-1)})/\sigma_{MH})}{\Phi((\dot{\gamma}_{j,k+1} - \dot{\gamma}_{j,k})/\sigma_{MH}) - \Phi((\gamma_{j,k-1}^{(t-1)} - \dot{\gamma}_{j,k})/\sigma_{MH}) }.
                \end{aligned}
                $}
            \end{equation}
Then set $\bm{\gamma}_{j}^{(t)} = \dot{\bm{\gamma}}_j$ with probability $AR$; if not accepted, $\bm{\gamma}_{j}^{(t)} =\bm{\gamma}_{j}^{(t-1)}$.
    \State Sample $Z^{(t)}_{ji}$ from $N(\mu_0 + f_{ji}^{(t)},1)I(\gamma^{(t)}_{j, Y_{ji}-1} < Z^{(t)}_{ji} \leq \gamma^{(t)}_{j, Y_{ji}})$.
    \State Set $t=t+1$ and repeat steps 2-4 $R^{ob}$ times to get the posterior distribution of the parameters.
    \Ensure $Z_{ji}$, $\bm{\gamma_j}$, $\{(\mathcal{T}, \mathcal{M})\}_{m=1}^M.$
\end{algorithmic}
\end{algorithm}

\section{Simulation}
\label{sec:simulation}

We conduct a simulation study under various scenarios to evaluate the performance of the proposed BML methods for constructing optimal DTRs with ordinal outcomes. 

\subsection{Simulation settings}
\label{sec:simulation1}

We consider 12 scenarios for generating the baseline variables $\bm{X}_{1}$, treatment variables $A_{1}$ and $A_{2}$, intermediate variables $\bm{X}_2$, and the latent continuous variable $Z_2$. We generate the stage-1 and stage-2 treatments based on $A_j \sim 2 \times \ber(0.5)-1, A_j \in \{-1,1 \}$, except for Sc.12, where $A_1|X_{11} \sim 2 \times \ber(\expit(0.2 X_{11}+0.5))-1$ and $A_2|X_{11}, X_{21} \sim 2 \times \ber(\expit(0.2 X_{11}+0.3 X_{21} + 0.5))-1$. Sc.12 is aimed to evaluate the model performance using data from observational studies where the treatment assignment is based on some confounders. To generate ordinal outcomes $Y_2$, we introduce a latent continuous variable $Z_{2}$ following a normal distribution. The details of the generating models for $Z_2$ are provided later. The ordinal outcome is derived by: $Y_{2i}=1$ if $Z_{2i} \leq -0.43$; $Y_i=2$ if $Z_{2i} \leq 0.43$; $Y_i=3$ if $Z_{2i} > 0.43$. These cutoff values are selected such that the baseline probability of falling into each outcome category is $1/3$. Here, ``baseline probability'' refers to the probability of being categorized into each outcome when the expected value of $Z_{2i}$ is zero. 

The first 9 scenarios are based on Murray et al. (2018) \citep{murray2018bayesian} and  Chakraborty et al. (2013) \citep{chakraborty2013inference} , where (i) $X_{11} \sim 2 \times \ber(0.5) -1$, (ii) $X_{21}|X_{11}, A_1 \sim 2 \times \ber(\expit(\delta_1 X_{11}+\delta_2 A_1))-1$, and (iii) $Z_2 = \beta_{20} + \beta_{21} X_{11} + \beta_{22} A_1 + \beta_{23} X_{11} A_1 + \beta_{24} X_{21} + \beta_{25} A_2 + \zeta_{21} A_2 X_{11} + \zeta_{22} A_2 A_1 + \zeta_{23} A_2 X_{21} + \epsilon$, $\epsilon \sim N(0,1)$ and $\expit(\cdot) = \frac{\exp(\cdot)}{1+\exp(\cdot)}$. The generating models in scenarios Sc.1 to Sc.9 are selected to represent varying degrees of non-regularity in terms of: 1) the probability of generating an individual with no stage-2 treatment effects and 2) the standardized effect size of the stage-2 treatment effects \citep{chakraborty2013inference}. In Sc.1 to Sc.9, we assume linear associations between the predictors and $Z_{2}$.  To provide a comprehensive evaluation, we include Sc.10, Sc.11, and Sc.12 with nonlinear terms: squared terms for Sc.10 and sine and cosine terms for Sc.11 and Sc.12, where (i) $X_{11} \sim N(0,1)$, (ii) $X_{21} \sim N(0,1)$, and (iii) for Sc.10, $Z_2 = \beta_{20} + \beta_{21} X_{11}^2 + \beta_{22} A_1 + \beta_{23} X_{11}^2 A_1 + \beta_{24} X^2_{21} + \beta_{25} A_2 + \zeta_{21} A_2 X^2_{21} + \epsilon$; for Sc.11 and Sc.12, $Z_2 = \beta_{20} + \beta_{21} \sin(X_{11}) + \beta_{22} A_1 + \beta_{23} \sin(X_{11}) A_1 + \beta_{24} \cos(X_{21}) + \beta_{25} A_2 + \zeta_{21} A_2 \cos(X_{21}) + \epsilon$.  Moreover, to evaluate the impact of noise variables on the model performance, we generate five noise variables $X_{12}, \dots, X_{16} \sim N(0,1)$ in addition to the variable $X_{11}$, which is the only baseline covariate related with the outcome. We incorporate all of these six variables when building models. The detailed parameter settings for these scenarios are presented in Table~\ref{tab1}. 

\begin{table}[htbp]
\centering
\caption{Parameters for scenarios Sc.1 to Sc.11. Here $\bm{\delta}_2$ is the vector of parameters for the intermediate variable $X_{21}$; $\bm{\beta}_2$ and $\bm{\zeta}_2$ are the vectors of parameters for the main effects and interaction effects in the generating models for $Z_2$.  `NR' represents `non-regular', `NNR' represents `near non-regular', and `RE' represents `regular'.}
\label{tab1}
\resizebox{\columnwidth}{!}{%
\begin{tabular}{lcccc}
\toprule
Scenario & $\bm{\delta}_2$ & $\bm{\beta}_2$ & $\bm{\zeta}_2$ & Type \\
\hline 
Sc.1 & $[0.50, 0.50]^T $ & $[0.00, 0.00, 0.00, 0.00, 0.00, 0.00]^T$ & 
$[0.00, 0.00, 0.00]^T$ & NR\\
Sc.2 & $[0.50, 0.50]^T $ & $[0.00, 0.00, 0.00, 0.00, 0.00, 0.10]^T$ & 
$[0.00, 0.00, 0.00]^T$& NNR\\
Sc.3 & $[0.50, 0.50]^T $ & $[0.00, 0.00, -0.50, 0.00, 0.00, 0.50]^T$ & $[0.00, 0.50, 0.00]^T$& NR\\
Sc.4 & $[0.50, 0.50]^T $ & $[0.00, 0.00, -0.50, 0.00, 0.00, 0.50]^T$ & $[0.00, 0.49, 0.00]^T$& NNR\\
Sc.5 & $[1.00, 0.00]^T $ & $[0.00, 0.00, -0.50, 0.00, 0.00, 1.00]^T$ & $[0.00, 0.50, 0.50]^T$& NR\\
Sc.6 & $[0.10, 0.10]^T $ & $[0.00, 0.00, -0.50, 0.00, 0.00, 0.25]^T$ & $[0.00, 0.50, 0.50]^T$& RE\\
Sc.7 & $[0.10, 0.10]^T $ & $[0.00, 0.00, -0.25, 0.00, 0.00, 0.75]^T$ & $[0.00, 0.50, 0.50]^T$& RE\\
Sc.8 & $[0.00, 0.00]^T $ & $[0.00, 0.00, 0.00, 0.00, 0.00, 0.25]^T$ & 
$[0.00, 0.25, 0.00]^T$& NR\\
Sc.9 & $[0.00, 0.00]^T $ & $[0.00, 0.00, 0.00, 0.00, 0.00, 0.25]^T$ & 
$[0.00, 0.24, 0.00]^T$& NNR\\
Sc.10 & NA  & $[0.00, 0.20, 0.20, -0.30, 0.10, -0.20]^T$ & 
$[0.20]^T$& Squared terms\\
Sc.11 & NA  & $[0.00, 0.40, -0.20, -1.00, 0.40, -0.70]^T$ & 
$[1.00]^T$& Sin and Cos terms\\
Sc.12 & NA  & $[0.00, 0.40, -0.20, -1.00, 0.40, -0.70]^T$ & 
$[1.00]^T$& Sin and Cos terms\\
\bottomrule
\end{tabular}
}
\end{table} 

We perform $100$ simulations. For each simulation, we first generate a training dataset of size $n_{tr} = 1000, 1500, 2000$ based on the assumed distributions. Subsequently, we build models using the training dataset and estimate the optimal DTRs for the testing dataset of size $n_{te}=1000$. All the stage-specific interactions between the available covariates and the treatment are included in the models. We compare the proposed BML approaches, including the BML with the Bayesian ordered probit model (BP), named BML-BP, and the BML with OBART, named BML-OBART, with Q-learning using the frequentist ordered probit model and the dWOLS method proposed by Jiang et al. \citep{jiang2024estimating}. For dWOLS, we use the inverse of the estimated propensity score as the final weights. The uncertainties of the estimated parameters can be obtained from the posterior distributions generated from the BML approaches. For Q-learning and dWOLS, we use the $m$-out-of-$n$ bootstrap to derive the confidence intervals for the parameters of interest \citep{chakraborty2013inference,simoneau2018non}. 

Following Li et al. (2024) \citep{li2024dynamic}, the evaluation metrics include: 1) bias at stage $j$: $Bias_j = \sum_{i=1}^{n_{te}} \{\hat{\psi}_{j}(\bm{H}_{ji})-\psi_{j}(\bm{H}_{ji})\} /n_{te}$, 2) mean squared error (MSE) at stage $j$: $MSE_j = \sum_{i=1}^{n_{te}} \{\hat{\psi}_{j}(\bm{H}_{ji})-\psi_{j}(\bm{H}_{ji})\}^2 /n_{te}$, and 3) the $95\%$ coverage rate (CR) at stage $j$: $CR_j = \sum_{i=1}^{n_{te}} I(\hat{\psi}^{lower}_{j}(\bm{H}_{ji}) \leq \psi_{j}(\bm{H}_{ji}) \leq \hat{\psi}^{upper}_{j}(\bm{H}_{ji}) ) / n_{te}$, where $\hat{\psi}^{lower}_{j}(\bm{H}_{ji})$ and $\hat{\psi}^{higher}_{j}(\bm{H}_{ji})$ are the lower and upper bounds of the confidence or credible intervals, respectively. These metrics assess the estimation of $\psi_{j}(\bm{H}_{ji})$, which determines the optimal decision rule (see Section~\ref{sec:methodsOBML1}). In addition, we compare the candidate approaches in terms of the stage-specific proportion of assigning the true optimal treatment \citep{chakraborty2016estimating}, denoted as 
$POT_j = \frac{1}{n_{te}}\sum_{i=1}^{n_{te} }I\{\hat{d}^{\ast}_j(\bm{H}_{ji}) =  d^{\ast}_j(\bm{H}_{ji})\}.$
To compare the estimated optimal DTR, the randomly assigned or observed treatment, and the true optimal DTR, we provide the probability that the outcome $Y_2=3$, denoted as $\sum_i^{n_{te}} I(Y_{2i}=3) / n_{te}$ under the estimated, true, and observed DTR in the testing dataset. These metrics are obtained by averaging over all the simulations. 

\subsection{Simulation results}
\label{sec:simulation2}

Table~\ref{tab2} shows the bias, the 95\% CRs, and MSEs, along with the proportion of assigning the true optimal treatment when the sample size of the training dataset is $n_{tr} = 1000$. The results for $n_{tr} = 1500, 2000$ show similar patterns and are provided in eTables 1 and 2 in Supplementary Materials. Under Sc.1 and Sc.2, all approaches generate nearly unbiased estimates of $\psi_j(\bm{H}_j)$ at both stages. In Sc.3 to Sc.9, BML-OBART outperforms or at least comparable to other approaches in terms of bias in most scenarios. However, the BML-OBART tends to generate overly conservative CRs (CRs higher than the nominal 95\% level) in most scenarios compared with BML-BP and Q-learning. Under Sc.10, Sc.11, and Sc. 12 with nonlinear terms in the true model, the outcome models specified by BML-BP, Q-learning, and dWOLS are misspecified; thus the CRs are lower compared with BML-OBART. Furthermore, the BML-OBART has the lowest MSEs in most scenarios at both stages, especially in Sc.10, Sc.11, and Sc.12. Frequentist methods dWOLS and Q-learning have higher MSE for stage-2 estimates compared with Bayesian methods. Moreover, BML-OBART outperforms other approaches in terms of the proportion of assigning the true optimal treatment ($POT$) to the participants in the test dataset at both stages in most scenarios, especially in Sc.10, Sc.11, and Sc.12. Q-learning, dWOLS, and BML-BP perform similarly at stage 2. 

\begin{table}[htbp]
\centering
\caption{The simulation results in terms of estimating $\psi_j(\bm{H}_j)$ with $n_{tr} = 1000$. POT denotes the proportion of assigning the true optimal treatment (POT) at each stage. }
\label{tab2}
\resizebox{0.9\columnwidth}{!}{%
\begin{tabular}[t]{llcccccccc}
\toprule
\multirow{2}{*}{Scenario} &  \multirow{2}{*}{Method} &\multicolumn{4}{c}{Stage 1} & \multicolumn{4}{c}{Stage 2} \\
&  & Bias & Cover & MSE & POT & Bias & Cover & MSE & POT\\
\midrule
Sc.1 NR & Q-learning & -0.006 & 0.994 & 0.041 & 1.000 & 0.000 & 0.950 & 0.048 & 1.000\\
 & dWOLS & -0.005 & 0.992 & 0.041 & 1.000 & 0.000 & 0.664 & 0.048 & 1.000\\
 & BML-BP & -0.001 & 1.000 & 0.003 & 1.000 & 0.000 & 1.000 & 0.004 & 1.000\\
 & BML-OBART & -0.005 & 1.000 & 0.007 & 1.000 & -0.001 & 0.994 & 0.009 & 1.000\\
Sc.2 NNR & Q-learning & -0.006 & 0.994 & 0.041 & 1.000 & 0.000 & 0.949 & 0.048 & 0.533\\
 & dWOLS & -0.004 & 0.991 & 0.042 & 1.000 & 0.000 & 0.667 & 0.049 & 0.532\\
 & BML-BP & -0.001 & 1.000 & 0.003 & 1.000 & -0.014 & 0.999 & 0.004 & 0.533\\
 & BML-OBART & -0.003 & 1.000 & 0.007 & 1.000 & 0.000 & 0.994 & 0.010 & 0.590\\
Sc.3 NR & Q-learning & -0.147 & 0.953 & 0.077 & 1.000 & 0.023 & 0.933 & 0.069 & 1.000\\
 & dWOLS & -0.146 & 0.949 & 0.078 & 1.000 & 0.023 & 0.634 & 0.070 & 1.000\\
 & BML-BP & -0.065 & 0.988 & 0.027 & 1.000 & -0.039 & 0.940 & 0.062 & 1.000\\
 & BML-OBART & -0.089 & 0.991 & 0.019 & 1.000 & 0.007 & 0.980 & 0.022 & 1.000\\
Sc.4 NNR & Q-learning & -0.138 & 0.957 & 0.074 & 0.769 & 0.022 & 0.932 & 0.069 & 0.758\\
 & dWOLS & -0.136 & 0.952 & 0.074 & 0.767 & 0.022 & 0.634 & 0.070 & 0.757\\
 & BML-BP & -0.052 & 0.990 & 0.025 & 0.713 & -0.039 & 0.939 & 0.061 & 0.758\\
 & BML-OBART & -0.082 & 0.995 & 0.019 & 0.826 & 0.006 & 0.977 & 0.022 & 0.858\\
Sc.5 NR & Q-learning & -0.023 & 0.973 & 0.063 & 0.652 & 0.061 & 0.929 & 0.097 & 1.000\\
 & dWOLS & -0.023 & 0.971 & 0.064 & 0.650 & 0.062 & 0.619 & 0.099 & 1.000\\
 & BML-BP & 0.019 & 0.988 & 0.034 & 0.611 & -0.019 & 0.941 & 0.080 & 1.000\\
 & BML-OBART & 0.005 & 0.998 & 0.014 & 0.727 & -0.016 & 0.960 & 0.043 & 1.000\\
Sc.6 RE & Q-learning & -0.020 & 0.970 & 0.056 & 0.998 & 0.005 & 0.944 & 0.065 & 0.991\\
 & dWOLS & -0.018 & 0.962 & 0.057 & 0.998 & 0.005 & 0.641 & 0.066 & 0.991\\
 & BML-BP & 0.097 & 0.961 & 0.053 & 0.998 & -0.024 & 0.947 & 0.058 & 0.991\\
 & BML-OBART & 0.002 & 0.993 & 0.015 & 1.000 & -0.006 & 0.964 & 0.031 & 0.999\\
Sc.7 RE & Q-learning & 0.040 & 0.957 & 0.067 & 0.825 & 0.033 & 0.940 & 0.072 & 0.995\\
 & dWOLS & 0.041 & 0.951 & 0.068 & 0.824 & 0.033 & 0.636 & 0.074 & 0.995\\
 & BML-BP & -0.031 & 0.982 & 0.032 & 0.832 & -0.035 & 0.942 & 0.066 & 0.995\\
 & BML-OBART & 0.049 & 0.990 & 0.017 & 0.979 & -0.004 & 0.961 & 0.034 & 0.998\\
Sc.8 NR & Q-learning & -0.115 & 0.967 & 0.066 & 0.958 & 0.007 & 0.941 & 0.053 & 1.000\\
 & dWOLS & -0.113 & 0.962 & 0.066 & 0.957 & 0.007 & 0.652 & 0.054 & 1.000\\
 & BML-BP & -0.212 & 0.892 & 0.073 & 0.968 & -0.088 & 0.910 & 0.053 & 1.000\\
 & BML-OBART & -0.073 & 1.000 & 0.016 & 1.000 & 0.007 & 0.988 & 0.017 & 1.000\\
Sc.9 NNR & Q-learning & -0.106 & 0.970 & 0.063 & 0.956 & 0.008 & 0.939 & 0.053 & 0.768\\
 & dWOLS & -0.104 & 0.965 & 0.063 & 0.955 & 0.008 & 0.648 & 0.054 & 0.769\\
 & BML-BP & -0.201 & 0.904 & 0.068 & 0.967 & -0.093 & 0.907 & 0.053 & 0.768\\
 & BML-OBART & -0.065 & 1.000 & 0.015 & 1.000 & 0.006 & 0.987 & 0.017 & 0.873\\
Sc.10 Squared & Q-learning & 0.022 & 0.592 & 0.643 & 0.442 & -0.068 & 0.626 & 0.509 & 0.535\\
 & dWOLS & 0.022 & 0.580 & 0.644 & 0.443 & -0.066 & 0.329 & 0.509 & 0.533\\
 & BML-BP & 0.100 & 0.357 & 0.610 & 0.442 & -0.039 & 0.336 & 0.460 & 0.534\\
 & BML-OBART & 0.021 & 0.834 & 0.341 & 0.799 & -0.065 & 0.819 & 0.298 & 0.856\\
Sc.11 Sin/Cos & Q-learning & -0.005 & 0.795 & 0.317 & 0.928 & 0.018 & 0.345 & 0.851 & 0.456\\
 & dWOLS & -0.004 & 0.774 & 0.317 & 0.928 & 0.018 & 0.166 & 0.853 & 0.456\\
 & BML-BP & 0.045 & 0.617 & 0.334 & 0.928 & 0.033 & 0.330 & 0.840 & 0.456\\
 & BML-OBART & -0.007 & 0.860 & 0.137 & 0.960 & 0.022 & 0.755 & 0.238 & 0.910\\
Sc.12 obs & Q-learning & 0.040 & 0.806 & 0.326 & 0.933 & 0.052 & 0.367 & 0.872 & 0.446\\
 & dWOLS & -0.005 & 0.821 & 0.328 & 0.929 & 0.022 & 0.164 & 0.873 & 0.440\\
 & BML-BP & 0.117 & 0.602 & 0.370 & 0.933 & 0.066 & 0.347 & 0.860 & 0.445\\
 & BML-OBART & 0.072 & 0.839 & 0.159 & 0.963 & 0.047 & 0.754 & 0.256 & 0.906\\
\bottomrule
\end{tabular}
}
\end{table}

To better understand the model's performance in estimating the contrast $\psi_j$ for each observation, we present boxplots of coverage rates for stage 1 and stage 2, as shown in eFigures 1 and 2. Each point represents the coverage rate for a specific individual across all simulation replicates. In most scenarios, the majority of individuals have desirable coverage rates. However, in Sc.10, Sc.11, and Sc.12, most individuals tend to have relatively low coverage rate in stage 1 or stage 2 when using Q-learning, dWOLS, or BML-BP. BML-OBART demonstrates improved performance in these scenarios. The `value' of the estimated DTR, the observed DTR, and the true optimal DTR are defined as the probability of getting the best outcome $Y=3$ under each respective DTR for the test dataset. We provide the results for comparing values using candidate approaches in eTable 3 in Supplementary Materials. In scenarios Sc.1 to Sc.9, all approaches perform similarly, with both the estimated and the true optimal decision rules yielding comparable results. However, in Sc. 10 to Sc. 12, the estimated optimal decision rules using BML-OBART are closer to the true optimal decision rules than other approaches. 

\begin{figure}[htbp]
\centerline{\includegraphics[width=0.7\linewidth]{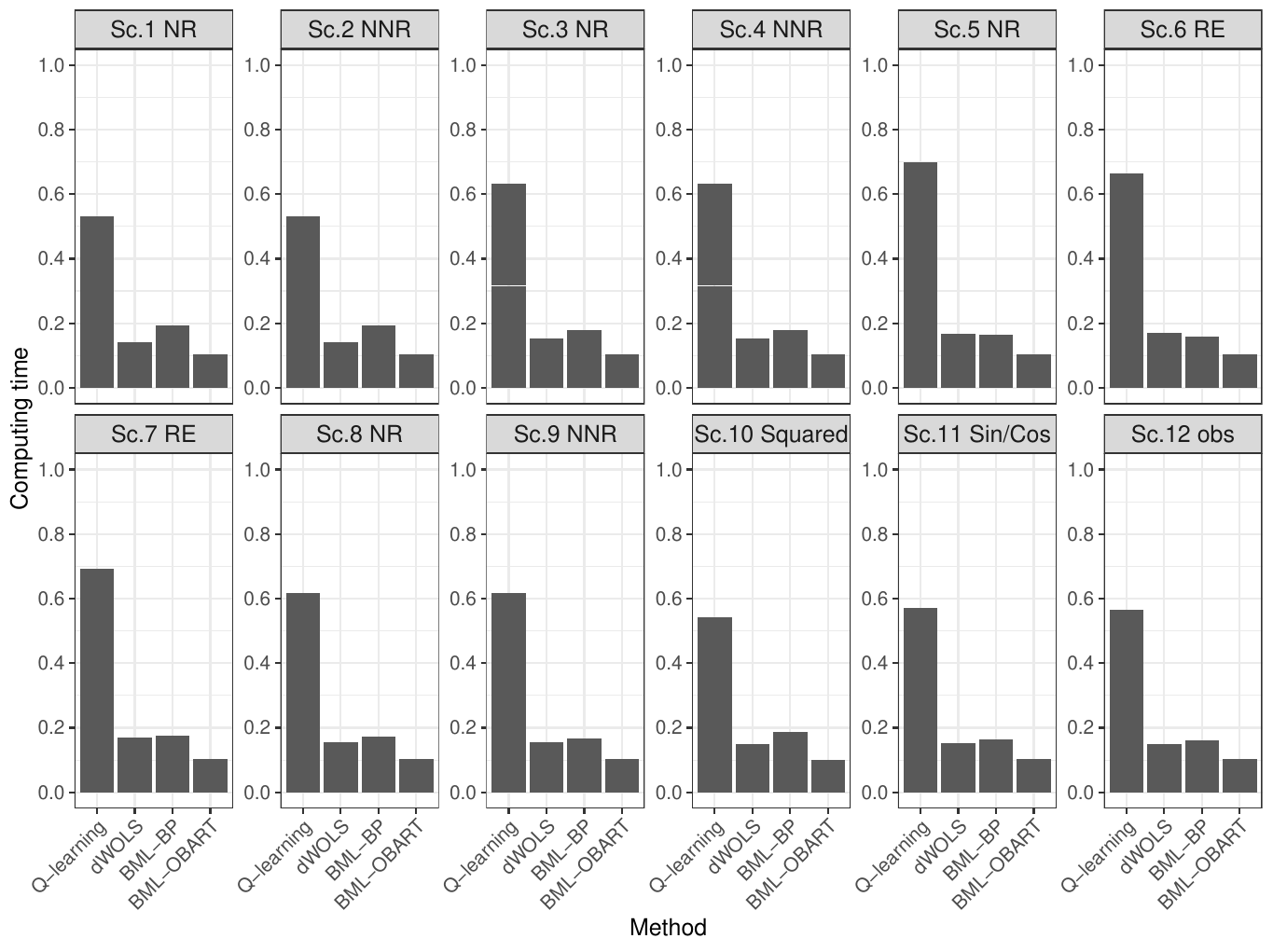}}
\caption{ The computing time for each candidate approach. The x-axis represents different approaches evaluated. Each panel corresponds to a scenario from Sc.1 to Sc.12. The y-axis represents the average running time for each candidate approach, with the unit `hour'. }
\label{simulation-time}
\end{figure}

Figure~\ref{simulation-time} compares the computing time for the candidate approaches. Q-learning requires more computing time than other approaches across all scenarios, followed by the BML-BP. One possible reason for Q-learning's relatively longer computation time is that it builds multiple ordered probit models with different pseudo outcomes as the response variable. The $m$-out-of-$n$ bootstrap steps will then repeat this procedure multiple times to get the confidence intervals of the model parameters. In contrast, the BIG sampler algorithm in BML approaches estimate stage-1 parameters directly from posterior distributions. Additionally, as we include all the noise variables and the interaction terms in the model, these noise variables may increase the computation time for the ordered probit model due to potential convergence difficulties. 

\section{Data Analysis for the smoking cessation trial}
\label{sec:application}

In this section, we demonstrate the proposed BML approaches using data from the smoking cessation trial introduced in Section~\ref{sec:intro}. The baseline covariates included gender, age, health maintenance organization (HMO) of the participant, number of cigarettes smoked per day (QuitCigsPerDay), education, race, motivation to quit (QuitOverallMotivBin), and self-efficacy (QuitOverallSEBin). For the purpose of demonstration, we convert continuous variables QuitOverallMotivBin and QuitOverallSEBin to binary and code the three-category race variables with two dummies, RaceWhite and RaceBlack. Although there were five two-level web-based intervention components in the Project Quit study, for the demonstration purpose we only consider the intervention component of providing the story of a hypothetical character who succeeded in quitting smoking, with higher and lower levels of the tailoring depth, coded as $A_1 = 1$ and $A_1 = -1$, respectively. 

\begin{table}[htbp]
\centering
\caption{Descriptive statistics (frequencies and percentages) for baseline and intermediate variables in the Project Quit and Forever Free studies. HMO: the health maintenance organizations; GHC: the Group Health Cooperative; HFHS: the Henry Ford Health System}
\label{tab3}
\resizebox{0.9\columnwidth}{!}{%
\begin{tabular}{lcc}
\hline
\textbf{Variables} & \textbf{Project Quit Sample (N = 1,153)} & \textbf{Forever Free Sample (N = 215)} \\ 
Gender ($X_{11}$) &  &  \\ 
\hspace{1em}Male ($X_{11} = 1$) & 465.0 (40.3\%) & 62.0 (28.8\%) \\ 
\hspace{1em}Female ($X_{11} = 0$) & 688.0 (59.7\%) & 153.0 (71.2\%) \\ 
Age ($X_{12}$) &  &  \\
\hspace{1em}Older than 50 ($X_{12} = 1$) & 470.0 (40.8\%) & 102.0 (47.4\%) \\
\hspace{1em}50 or younger ($X_{12} = 0$) & 683.0 (59.2\%) & 113.0 (52.6\%) \\ 
HMO ($X_{13}$) &  &  \\
\hspace{1em}GHC ($X_{13} = 1$) & 644.0 (55.9\%) & 113.0 (52.6\%) \\
\hspace{1em}HFHS ($X_{13} = 0$) & 509.0 (44.1\%) & 102.0 (47.4\%) \\ 
QuitCigsPerDay ($X_{14}$) &  &  \\
\hspace{1em}More than 20/day ($X_{14} = 1$) & 354.0 (30.7\%) & 59.0 (27.4\%) \\
\hspace{1em}20 or fewer/day ($X_{14} = 0$) & 799.0 (69.3\%) & 156.0 (72.6\%) \\ 
Education ($X_{15}$) &  &  \\
\hspace{1em}$>$ High school ($X_{15} = 1$) & 740.0 (64.2\%) & 157.0 (73.0\%) \\
\hspace{1em}$\leq$ High school ($X_{15} = 0$) & 413.0 (35.8\%) & 58.0 (27.0\%) \\ 
RaceWhite ($X_{16}$) &  &  \\
\hspace{1em}White ($X_{16} = 1$) & 910.0 (78.9\%) & 166.0 (77.2\%) \\
\hspace{1em}Non-White ($X_{16} = 0$) & 243.0 (21.1\%) & 49.0 (22.8\%) \\ 
RaceBlack ($X_{17}$) &  &  \\
\hspace{1em}Black ($X_{17} = 1$) & 133.0 (11.5\%) & 27.0 (12.6\%) \\
\hspace{1em}Non-Black ($X_{17} = 0$) & 1,020.0 (88.5\%) & 188.0 (87.4\%) \\ 
QuitOverallMotivBin ($X_{18}$) &  &  \\
\hspace{1em}High motivation ($X_{18} = 1$) & 562.0 (48.7\%) & 101.0 (47.0\%) \\
\hspace{1em}Low motivation ($X_{18} = 0$) & 591.0 (51.3\%) & 114.0 (53.0\%) \\ 
QuitOverallSEBin ($X_{19}$) &  &  \\
\hspace{1em}High self-efficacy ($X_{19} = 1$) & 634.0 (55.0\%) & 115.0 (53.5\%) \\
\hspace{1em}Low self-efficacy ($X_{19} = 0$) & 519.0 (45.0\%) & 100.0 (46.5\%) \\ 
Tailoring depth ($A_1$) &  &  \\
\hspace{1em}High tailoring ($A_1 = 1$) & 604.0 (52.4\%) & 110.0 (51.2\%) \\
\hspace{1em}Low tailoring ($A_1 = -1$) & 549.0 (47.6\%) & 105.0 (48.8\%) \\ 
PQ6Quitstatus ($X_{21}$) &  &  \\
\hspace{1em}Quit ($X_{21} = 1$) & 370.0 (32.1\%) & 47.0 (21.9\%) \\
\hspace{1em}Not quit ($X_{21} = 0$) & 783.0 (67.9\%) & 168.0 (78.1\%) \\ 
PQ6OverallMotivBin ($X_{22}$) &  &  \\
\hspace{1em}High motivation ($X_{22} = 1$) & 605.0 (52.5\%) & 114.0 (53.0\%) \\
\hspace{1em}Low motivation ($X_{22} = 0$) & 548.0 (47.5\%) & 101.0 (47.0\%) \\ 
PQ6OverallSEBin ($X_{23}$) &  &  \\
\hspace{1em}High self-efficacy ($X_{23} = 1$) & 674.0 (58.5\%) & 121.0 (56.3\%) \\
\hspace{1em}Low self-efficacy ($X_{23} = 0$) & 479.0 (41.5\%) & 94.0 (43.7\%) \\ 
PQ6OverallSatBin ($X_{24}$) &  &  \\
\hspace{1em}High satisfaction ($X_{24} = 1$) & 656.0 (56.9\%) & 123.0 (57.2\%) \\
\hspace{1em}Low satisfaction ($X_{24} = 0$) & 497.0 (43.1\%) & 92.0 (42.8\%) \\ 
Forever Free Intervention ($A_2$) &  &  \\
\hspace{1em}Intervention ($A_2 = 1$) & 145.0 (67.4\%) & 145.0 (67.4\%) \\
\hspace{1em}Control ($A_2 = -1$) & 70.0 (32.6\%) & 70.0 (32.6\%) \\ 

\hline
\end{tabular}%
}
\end{table}

\begin{figure}[htbp]
    \centering
     \renewcommand{\thesubfigure}{\Alph{subfigure}}
     
     \captionsetup[subfigure]{labelformat=simple, labelsep=period, font=large} 
     
    \subfloat[]{%
        \includegraphics[width=0.5\linewidth, trim=25 70 25 70, clip]{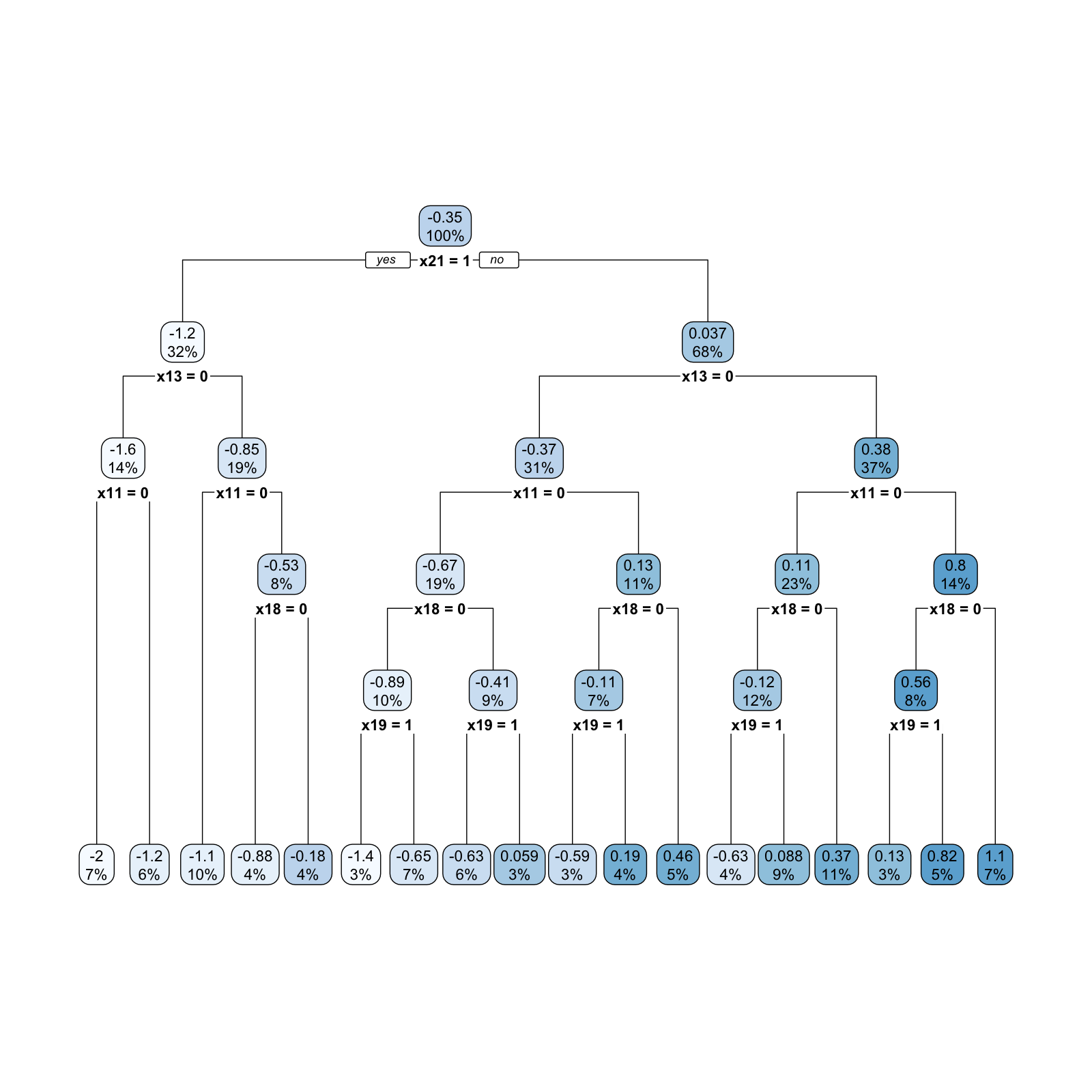}%
        }%
    \hfill%
    \subfloat[]{%
        \includegraphics[width=0.5\linewidth, trim=25 70 25 70, clip]{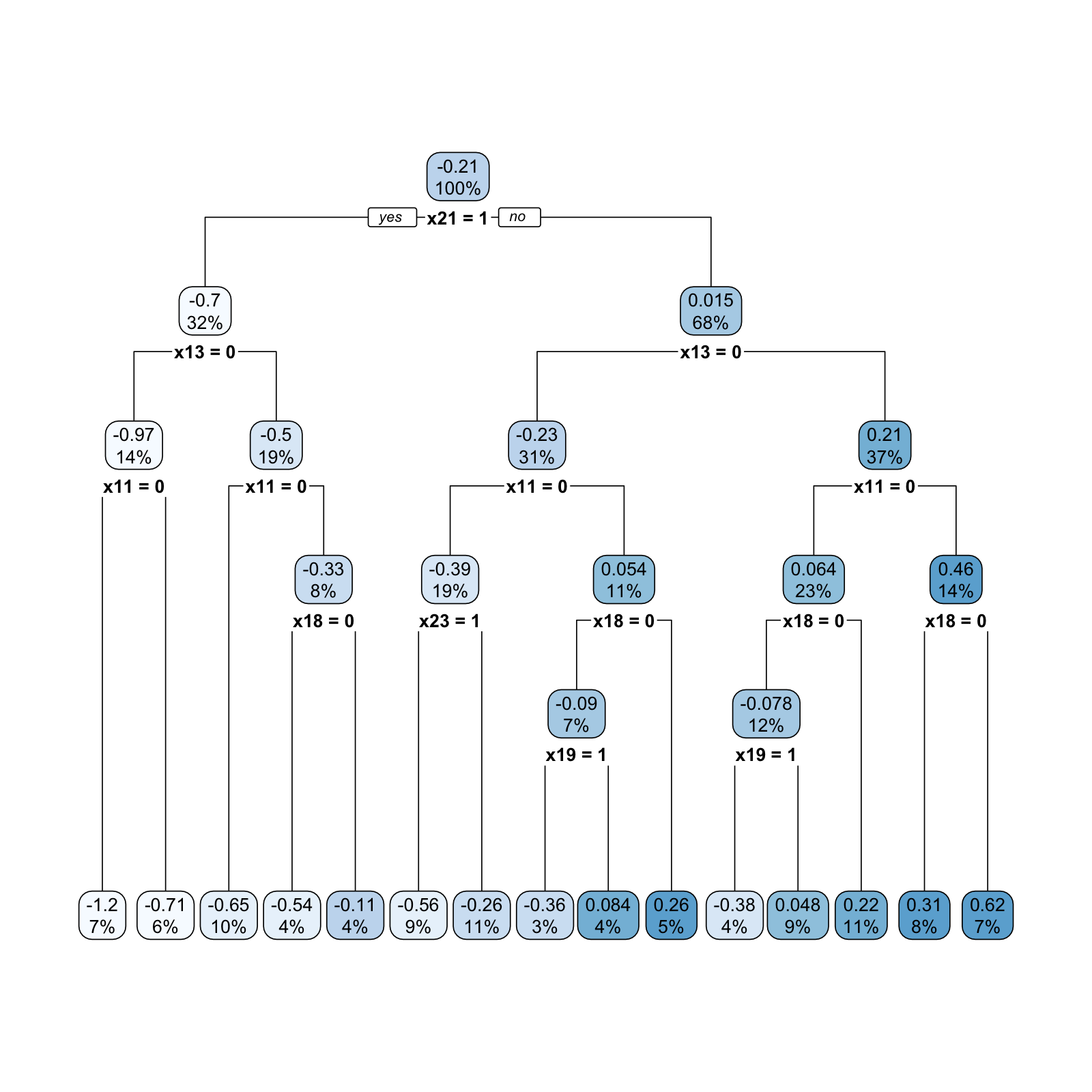}%
        }%
    \subfloat[]{%
        \includegraphics[width=0.5\linewidth, trim=25 70 25 70, clip]{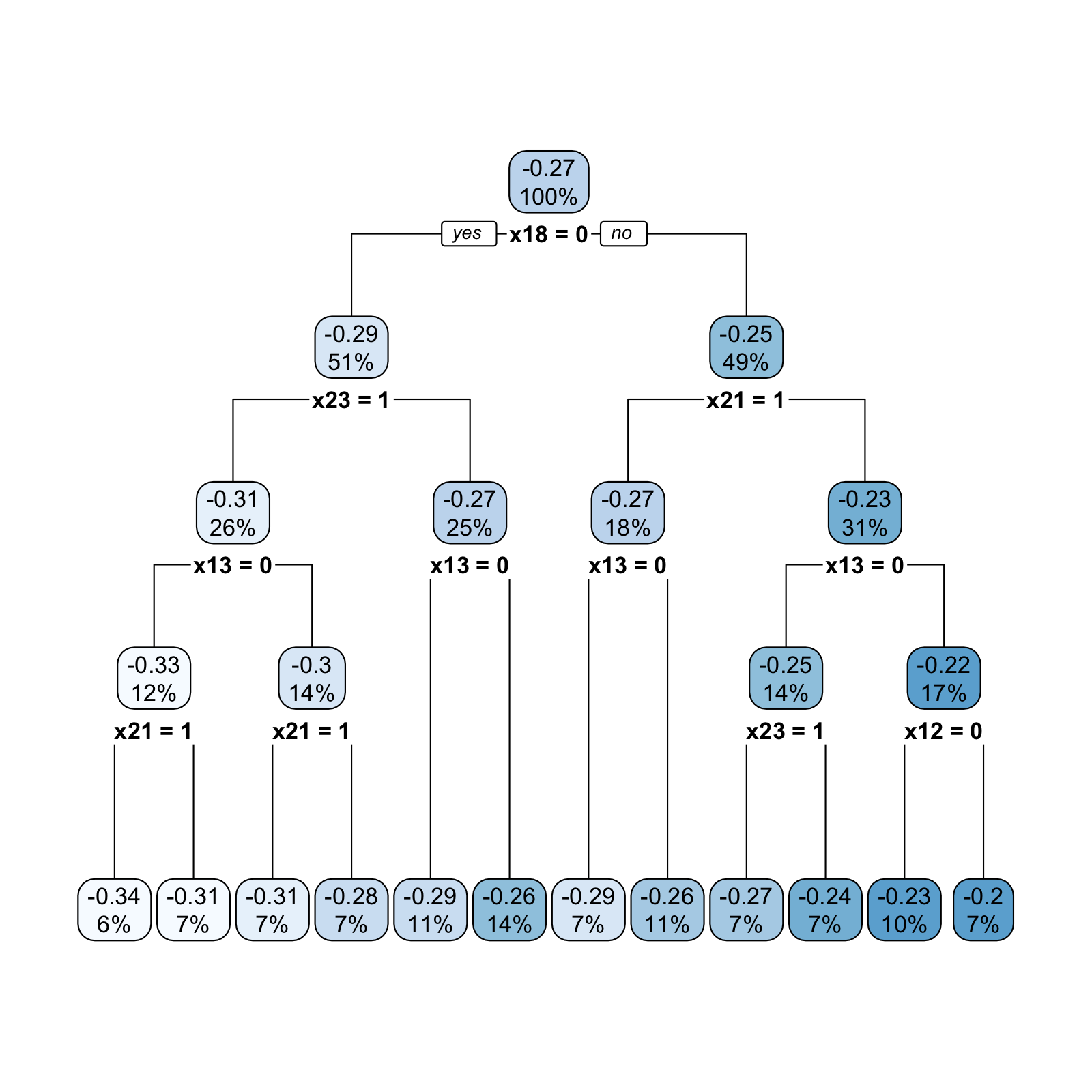}%
        }%
    \caption{Application results for constructing optimal stage-2 decision rule using `fit-the-fit' approach. A) Q-learning, B) the BML approach with the ordered probit model, BML-BP, and C) the BML approach with the OBART model, BML-OBART. }
    \label{appfig1}
\end{figure}

\begin{figure}[htbp]
    \centering
     \renewcommand{\thesubfigure}{\Alph{subfigure}}
    \subfloat[]{%
        \includegraphics[width=0.5\linewidth, trim=25 30 25 30, clip]{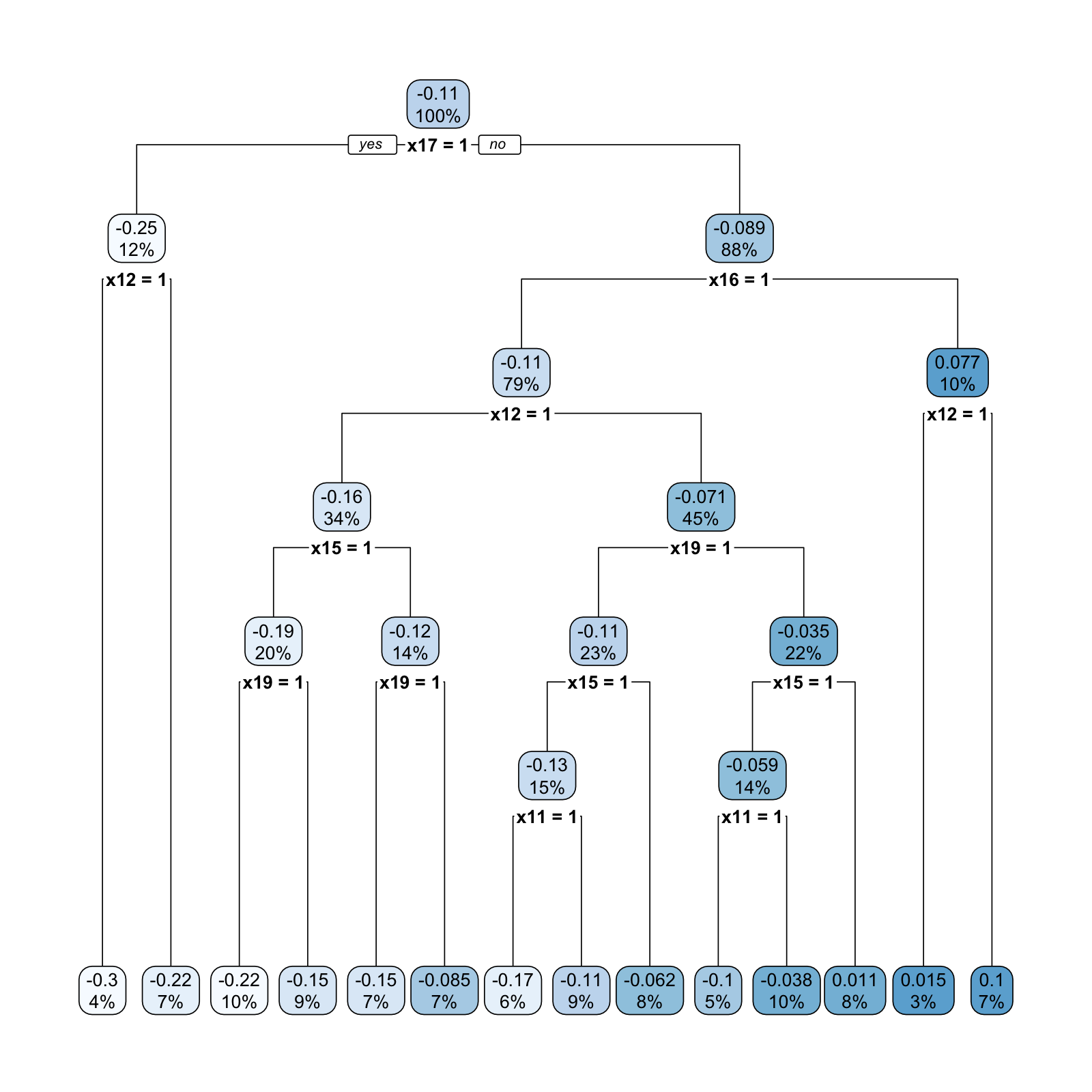}%
        }%
    \hfill%
    \subfloat[]{%
        \includegraphics[width=0.5\linewidth, trim=25 70 25 70, clip]{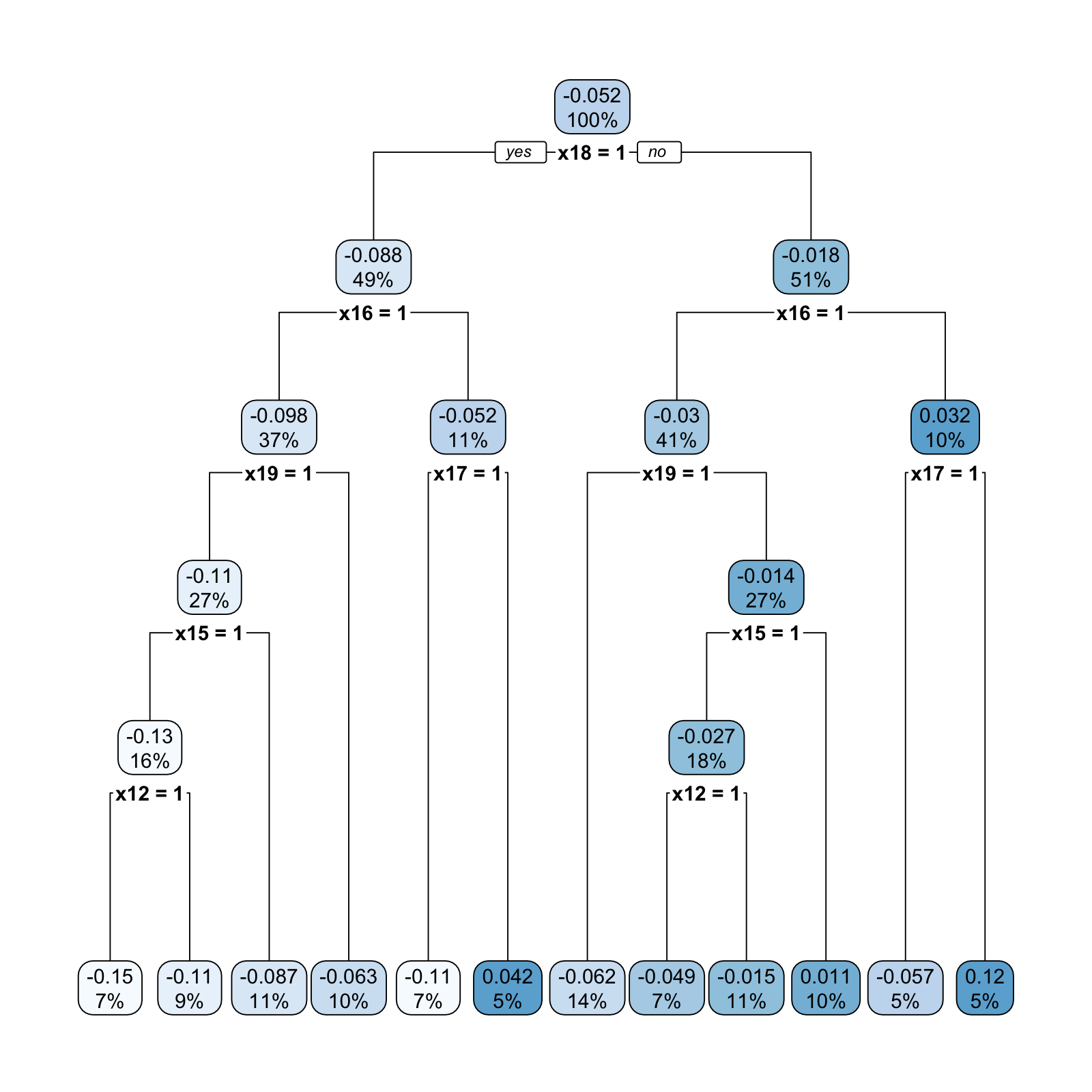}%
        }%
    \subfloat[]{%
        \includegraphics[width=0.5\linewidth, trim=25 50 25 50, clip]{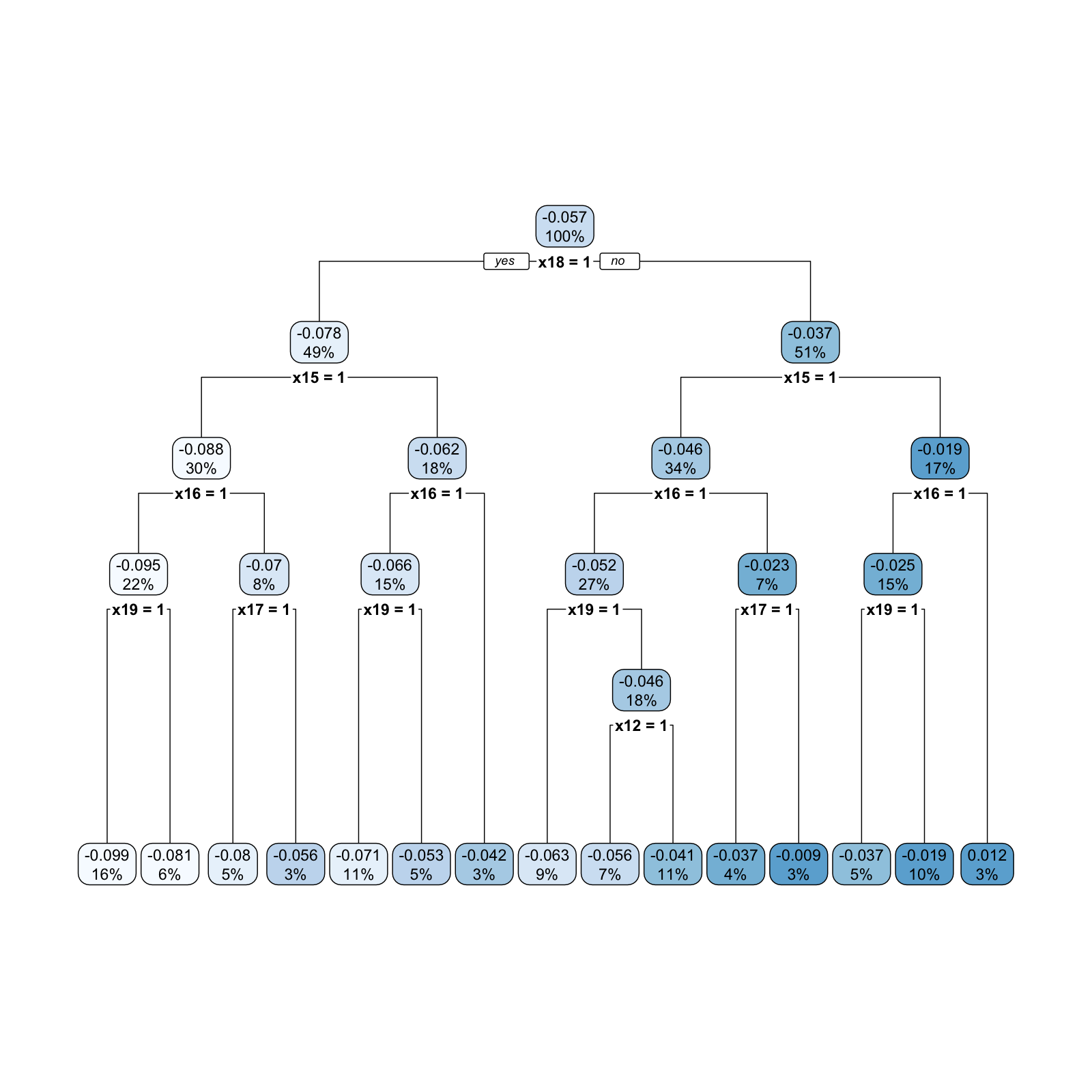}%
        }%
    \caption{Application results for constructing optimal stage-1 decision rule using `fit-the-fit' approach. A) Q-learning, B) the BML approach with the ordered probit model, BML-BP, and C) the BML approach with the OBART model, BML-OBART. }
    \label{appfig2}
\end{figure}

The participants were asked to complete a computer-assisted telephone interview six months after the initial intervention. The intermediate tailoring variables collected during this interview include the quit status based on the last seven days (PQ6Quitstatus), self-efficacy (PQ6OverallSEBin), the motivation to quit (PQ6OverallMotivBin), and the level of satisfaction (PQ6OverallSatBin). These covariates were transformed in the same way as the baseline covariates. Additionally, participants were asked if they wanted to continue to participate in the Forever Free study. Those who agreed were randomized into one of the four intervention arms or the control arm. The intervention arms provided smoking relapse prevention booklets with different levels of tailoring content, while the control arm did not provide any of the booklets and only gave an encouraging message to help them quit smoking. For analysis, we combined the four intervention arms into a single group, coding it as $A_{2} = 1$ while coding the control arm as $A_2 = -1$. The detailed descriptive statistics about the baseline variables and the interventions in the Project Quit (PQ) study and the Forever Free (FF) study are provided in Table~\ref{tab3}.

To provide a more comprehensive assessment based on the long-term smoking cessation, we use the number of months not smoked by a participant ranging from 0 to 6 months. As introduced in Section~\ref{sec:intro}, this variable is collaposed into an ordinal outcome with three categories: 0 for participants who smoked throughout, 1 for those who abstained for more than 0 but less than or equal to 3 months, and 2 for those who abstained for more than 3 months but less than or equal to 6 months. This re-coded ordinal outcome, $Y_2$, serves as the primary outcome to construct optimal decision rules for aiding smoking cessation in the following analysis. In our analysis, we include only those who consented to participate in the Forever Free study and for whom the number of months not smoked ($Y_2$) was collected. This subgroup constitutes the Forever Free sample for constructing optimal stage-2 decision rules. For stage-1 estimation, in addition to the Forever Free sample, we also include those who did not consent to participate in the Forever Free study and for whom  $X_{21}$, the quit status after stage 1 randomization was collected for constructing stage-1 optimal decision rules. After removing ineligible participants, we observe a maximum missing rate of 0.95\% among baseline and intermediate variables. Given the negligible missing rates and the fact that imputation of missing values is not the focus of this paper, we employ mean imputation for continuous variables and mode imputation for binary variables. After preprocessing the data, there are 1,153 participants in the Project Quit study (stage 1), with 215 participants consenting to participate in the Forever Free study (stage 2). 

We employ Q-learning, BML-BP, and BML-OBART to estimate $\psi_j(\bm{H}_j)$ for each individual. To get the confidence interval of the contrast $\psi_j(\bm{H}_j)$ for Q-learning, we use the $m$-out-of-$n$ bootstrap with tuning parameter $\alpha=0.05$ and $1,000$ bootstrap samples. For BML-BP and BML-OBART, we draw $4,000$ random samples from the posterior distributions of the parameters of interest with $2,000$ burn-in draws. Note that BML approaches only provide estimated contrast parameters and the corresponding credible intervals, from which the optimal decision rules are derived. To visualize the estimated optimal decision rule at each stage and make it more interpretable, we adopt the `fit-the-fit' approach \citep{logan2019decision} to generate a single tree structure decision rule for each stage using the \textit{rpart} package \citep{rpact2024}. The posterior mean of $\psi_j(\bm{H}_j)$ is used as the outcome for building the single regression tree. Figure~\ref{appfig1} and Figure~\ref{appfig2} illustrate the tree structure decision rules at stages 2 and 1, respectively. We specify the minimum number of observations that must exist in a node for a split and the minimum number of observations in any terminal node as 50 and 17, respectively. The initial complexity parameter is specified to be the default value 0.01 for both stages 1 and 2. 

For the stage 2 decision rule shown in Figure~\ref{appfig1}, HMO, QuitOverallMotivBin, and PQ6Quitstatus are identified by all three approaches. Another finding is that both Q-learning and BML-BP detect heterogeneity in treatment effects at stage 2, whereas BML-OBART indicates that all participants benefit from the control. For the stage 1 decision rule in Figure~\ref{appfig2}, Age, Education, RaceWhite, RaceBlack, and QuitOverallSEBin are selected by all these three approaches. All these three approaches indicate that the majority of the smokers benefit from the low tailoring depth intervention. The values of $R^2$ goodness of fit for the tree structures in Figure~\ref{appfig1} and Figure~\ref{appfig2} are around $0.8$ for all approaches, indicating that the tree model provides results comparable to those of the original models by these approaches. In addition, eFigures 3 and 4 in Supplementary Materials show the distribution of $\psi_j, j=2, 1$ for each individual in the testing dataset, ordered from the smallest to the largest. The intervals generated by all these approaches include zero for most of the individuals, indicating insufficient evidence of treatment effects for all individuals at both stages. BML-OBART generates narrower intervals than Q-learning and BML-BP, which aligns with the findings of Li et al. (2024) \citep{li2024dynamic}. The overly conservative intervals of Q-learning and BML-BP may result from overfitting when all covariates and their interactions with the treatment are included in the model. 

\section{Discussion}
\label{sec:discussion}

Ordinal outcomes are frequently used as one of the primary endpoints in clinical trials. However, there is limited research on constructing optimal DTRs with ordinal outcomes. In this paper, we extend the BML approach to estimate optimal DTRs with ordinal outcomes. To address nonlinear associations in the outcome model, we propose the novel OBART model, allowing BART to handle ordinal outcomes. We then incorporate the OBART model into the BML framework to estimate the parameters of interest and quantify the uncertainties. We conduct extensive simulation studies under various scenarios to evaluate model performance. The results demonstrate that the BML-OBART outperforms the competing methods in terms of Bias, MSE, and the proportion of assigning true optimal reatments when there are nonlinear associations between predictors and outcomes. However, BML-OBART tends to generate overly conservative confidence intervals compared with other approaches in some scenarios. We have illustrated the application of our proposed approach using a smoking cessation trial data. All three methods (Q-leaning, BML-BP, and BML-OBART) show insufficient evidence to declare significant treatment effects at both stages, except for Q-learning, which identifies significant treatment effects for a small number of participants at both stages. 

The primary limitation of the proposed method is that Algorithm~\ref{algorithm_obart} is fitted $R^{bml}$ times at the step 4 in Algorithm~\ref{algorithm_bml-bp}, which can be computationally intensive. Employing parallel computation can significantly reduce this computational burden. Additionally, we use default priors of the tree parameters and uniform priors for cutoff values in the BML approaches. Informative priors can be incorporated based on Johnson and Albert (1999) \citep{Johnson1999} if investigators have prior information about the frequency of each outcome category. Model performance may be further improved by choosing the optimal hyperparameters using cross validation. Although treating ordinal outcomes as continuous and then applying the original BML approach is a potential alternative, we do not consider it in our simulation study due to differences in the definitions of ``optimal decision rules'' compared to our proposed method. Specifically, in Section~\ref{sec:methodsOBML}, the optimal decision rules aim to maximize the latent continuous variables, thereby increasing the probability of receiving the best outcome. However, the continuous BML approach seeks to maximize the ordinal outcome under the assumption of normality. When the number of categories is small, this assumption may be unrealistic and lead to uninterpretable results. Another limitation of the proposed BML approach is the overly conservative credible intervals. Future work may explore alternative method to generate credible intervals closer to the nominal level. 

\section*{Acknowledgments}
X. Wang is supported by a PhD scholarship from the Duke-NUS Medical School, Singapore. Dr. B. Chakraborty would like to acknowledge support from the grant MOE-T2EP20122-0013 from the Ministry of Education, Singapore. 

\section*{DATA AVAILABILITY}

The data used in this study are available upon reasonable request to the corresponding author at \href{bibhas.chakraborty@duke-nus.edu.sg}{bibhas.chakraborty@duke-nus.edu.sg}. The \textit{OBART} package can be accessed via \url{https://github.com/cherylwaal/OBART} and the example code for implementing BML-OBART and BML-BP are available at \url{https://github.com/cherylwaal/BMLordinal}. 

\section*{Conflict of interest}

The authors declare no potential conflict of interests.

\bibliography{ms}

\section*{Supporting information}

Additional supporting information may be found in the online version of the article at the publisher’s website.

\end{document}


\maketitle

\section{Simulation results}

\begin{figure}[H]
\centerline{\includegraphics[width=0.65\linewidth]{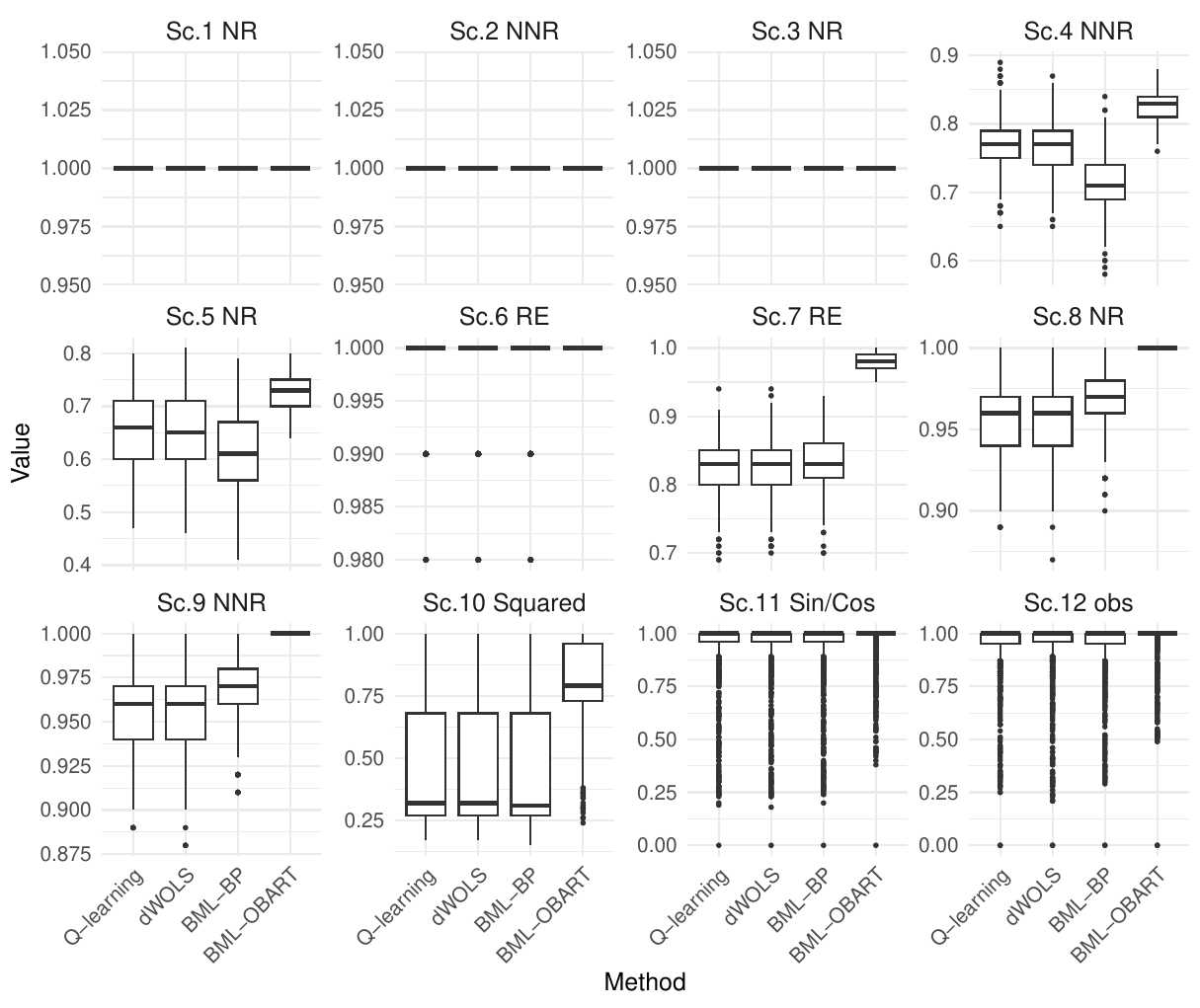}}
\caption{The boxplot for the coverage rate of the 95\% confidence or credible intervals of $\psi_1$. Each point represents the coverage rate for one observation in the test dataset. }
\end{figure}

\begin{figure}[H]
\centerline{\includegraphics[width=0.65\linewidth]{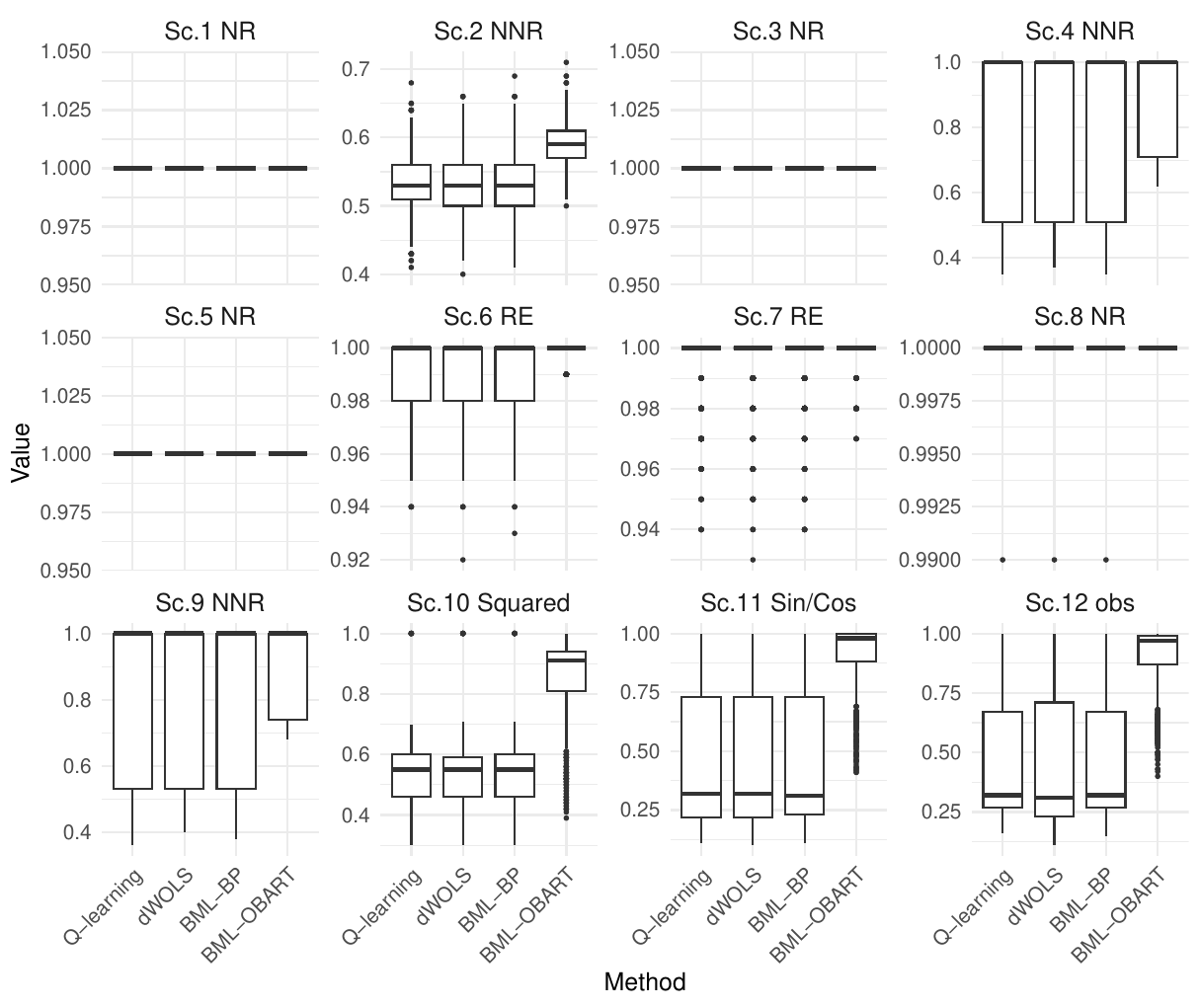}}
\caption{The boxplot for the coverage rate of the 95\% confidence or credible intervals of $\psi_2$. Each point represents the coverage rate for one observation in the test dataset. }
\end{figure}

\begin{table}[H]
\centering
\caption{The simulation results in terms of estimating $\psi_j(\bm{H}_j)$ with $n_{tr} = 1500$. POT denotes the proportion of assigning the true optimal treatment (POT) at each stage. }
\label{tab1}
\resizebox{0.8\columnwidth}{!}{%
\begin{tabular}[t]{llcccccccc}
\toprule
\multirow{2}{*}{Scenario} &  \multirow{2}{*}{Method} &\multicolumn{4}{c}{Stage 1} & \multicolumn{4}{c}{Stage 2} \\
&  & Bias & Cover & MSE & POT & Bias & Cover & MSE & POT\\
\midrule
Sc.1 NR & Q-learning & -0.008 & 0.991 & 0.029 & 1.000 & 0.003 & 0.939 & 0.034 & 1.000\\
 & dWOLS & -0.008 & 0.989 & 0.029 & 1.000 & 0.003 & 0.652 & 0.034 & 1.000\\
 & BML-BP & -0.003 & 1.000 & 0.003 & 1.000 & 0.000 & 0.998 & 0.003 & 1.000\\
 & BML-OBART & -0.009 & 0.998 & 0.006 & 1.000 & 0.004 & 0.989 & 0.008 & 1.000\\
Sc.2 NNR & Q-learning & -0.008 & 0.990 & 0.030 & 1.000 & 0.003 & 0.940 & 0.034 & 0.550\\
 & dWOLS & -0.008 & 0.987 & 0.030 & 1.000 & 0.004 & 0.654 & 0.034 & 0.551\\
 & BML-BP & -0.003 & 1.000 & 0.003 & 1.000 & -0.014 & 0.997 & 0.004 & 0.551\\
 & BML-OBART & -0.009 & 0.999 & 0.006 & 1.000 & 0.004 & 0.992 & 0.008 & 0.610\\
Sc.3 NR & Q-learning & -0.114 & 0.943 & 0.052 & 1.000 & 0.025 & 0.945 & 0.041 & 1.000\\
 & dWOLS & -0.114 & 0.940 & 0.052 & 1.000 & 0.027 & 0.654 & 0.041 & 1.000\\
 & BML-BP & -0.051 & 0.986 & 0.019 & 1.000 & -0.016 & 0.945 & 0.037 & 1.000\\
 & BML-OBART & -0.070 & 0.995 & 0.013 & 1.000 & 0.020 & 0.968 & 0.015 & 1.000\\
Sc.4 NNR & Q-learning & -0.105 & 0.947 & 0.050 & 0.748 & 0.025 & 0.946 & 0.041 & 0.782\\
 & dWOLS & -0.105 & 0.944 & 0.050 & 0.748 & 0.026 & 0.650 & 0.041 & 0.782\\
 & BML-BP & -0.038 & 0.989 & 0.018 & 0.691 & -0.016 & 0.946 & 0.037 & 0.783\\
 & BML-OBART & -0.063 & 0.998 & 0.012 & 0.810 & 0.019 & 0.973 & 0.015 & 0.880\\
Sc.5 NR & Q-learning & -0.002 & 0.965 & 0.042 & 0.638 & 0.041 & 0.939 & 0.057 & 1.000\\
 & dWOLS & -0.002 & 0.964 & 0.042 & 0.639 & 0.043 & 0.641 & 0.058 & 1.000\\
 & BML-BP & 0.029 & 0.982 & 0.028 & 0.595 & -0.011 & 0.944 & 0.051 & 1.000\\
 & BML-OBART & 0.020 & 0.998 & 0.011 & 0.667 & -0.002 & 0.956 & 0.031 & 1.000\\
Sc.6 RE & Q-learning & -0.005 & 0.952 & 0.041 & 0.999 & 0.014 & 0.936 & 0.044 & 0.996\\
 & dWOLS & -0.004 & 0.943 & 0.041 & 0.999 & 0.015 & 0.637 & 0.045 & 0.996\\
 & BML-BP & 0.074 & 0.951 & 0.040 & 0.999 & -0.006 & 0.936 & 0.041 & 0.996\\
 & BML-OBART & 0.003 & 0.996 & 0.010 & 1.000 & 0.009 & 0.963 & 0.021 & 1.000\\
Sc.7 RE & Q-learning & 0.040 & 0.949 & 0.045 & 0.864 & 0.031 & 0.930 & 0.050 & 0.997\\
 & dWOLS & 0.039 & 0.943 & 0.045 & 0.863 & 0.032 & 0.629 & 0.050 & 0.997\\
 & BML-BP & -0.024 & 0.977 & 0.023 & 0.870 & -0.014 & 0.932 & 0.046 & 0.997\\
 & BML-OBART & 0.045 & 0.993 & 0.012 & 0.992 & 0.012 & 0.954 & 0.026 & 0.999\\
Sc.8 NR & Q-learning & -0.095 & 0.949 & 0.046 & 0.979 & 0.007 & 0.944 & 0.034 & 1.000\\
 & dWOLS & -0.095 & 0.945 & 0.046 & 0.979 & 0.007 & 0.650 & 0.035 & 1.000\\
 & BML-BP & -0.165 & 0.894 & 0.051 & 0.984 & -0.060 & 0.923 & 0.034 & 1.000\\
 & BML-OBART & -0.060 & 0.991 & 0.013 & 0.998 & 0.009 & 0.969 & 0.014 & 1.000\\
Sc.9 NNR & Q-learning & -0.087 & 0.953 & 0.044 & 0.976 & 0.007 & 0.945 & 0.034 & 0.773\\
 & dWOLS & -0.086 & 0.950 & 0.044 & 0.976 & 0.008 & 0.650 & 0.035 & 0.774\\
 & BML-BP & -0.157 & 0.904 & 0.048 & 0.982 & -0.064 & 0.920 & 0.034 & 0.773\\
 & BML-OBART & -0.051 & 0.991 & 0.012 & 0.999 & 0.008 & 0.975 & 0.013 & 0.866\\
Sc.10 Squared & Q-learning & 0.017 & 0.475 & 0.621 & 0.429 & -0.057 & 0.527 & 0.488 & 0.533\\
 & dWOLS & 0.016 & 0.465 & 0.621 & 0.431 & -0.057 & 0.255 & 0.488 & 0.533\\
 & BML-BP & 0.097 & 0.280 & 0.607 & 0.429 & -0.038 & 0.258 & 0.458 & 0.533\\
 & BML-OBART & 0.011 & 0.869 & 0.252 & 0.901 & -0.049 & 0.831 & 0.233 & 0.911\\
Sc.11 Sin/Cos & Q-learning & -0.015 & 0.658 & 0.300 & 0.929 & 0.026 & 0.269 & 0.829 & 0.448\\
 & dWOLS & -0.014 & 0.630 & 0.300 & 0.929 & 0.026 & 0.128 & 0.829 & 0.449\\
 & BML-BP & 0.019 & 0.510 & 0.316 & 0.929 & 0.036 & 0.260 & 0.824 & 0.448\\
 & BML-OBART & -0.008 & 0.866 & 0.109 & 0.963 & 0.021 & 0.794 & 0.159 & 0.916\\
Sc.12 obs & Q-learning & 0.064 & 0.681 & 0.306 & 0.943 & 0.046 & 0.282 & 0.851 & 0.434\\
 & dWOLS & 0.026 & 0.706 & 0.303 & 0.940 & 0.021 & 0.127 & 0.854 & 0.431\\
 & BML-BP & 0.116 & 0.512 & 0.342 & 0.944 & 0.056 & 0.271 & 0.846 & 0.434\\
 & BML-OBART & 0.066 & 0.875 & 0.114 & 0.961 & 0.029 & 0.770 & 0.173 & 0.924\\
\bottomrule
\end{tabular}
}
\end{table}

\begin{table}[H]
\centering
\caption{The simulation results in terms of estimating $\psi_j(\bm{H}_j)$ with $n_{tr} = 2000$. POT denotes the proportion of assigning the true optimal treatment (POT) at each stage. }
\label{tab2}
\resizebox{0.8\columnwidth}{!}{%
\begin{tabular}[t]{llcccccccc}
\toprule
\multirow{2}{*}{Scenario} &  \multirow{2}{*}{Method} &\multicolumn{4}{c}{Stage 1} & \multicolumn{4}{c}{Stage 2} \\
&  & Bias & Cover & MSE & POT & Bias & Cover & MSE & POT\\
\midrule
Sc.1 NR & Q-learning & -0.018 & 0.992 & 0.021 & 1.000 & -0.002 & 0.949 & 0.023 & 1.000\\
 & dWOLS & -0.019 & 0.989 & 0.021 & 1.000 & -0.002 & 0.679 & 0.023 & 1.000\\
 & BML-BP & -0.006 & 1.000 & 0.002 & 1.000 & 0.000 & 0.998 & 0.002 & 1.000\\
 & BML-OBART & -0.016 & 1.000 & 0.005 & 1.000 & -0.003 & 0.988 & 0.005 & 1.000\\
Sc.2 NNR & Q-learning & -0.018 & 0.992 & 0.021 & 1.000 & -0.001 & 0.949 & 0.023 & 0.545\\
 & dWOLS & -0.018 & 0.989 & 0.021 & 1.000 & -0.001 & 0.681 & 0.023 & 0.546\\
 & BML-BP & -0.006 & 1.000 & 0.002 & 1.000 & -0.014 & 0.997 & 0.002 & 0.545\\
 & BML-OBART & -0.016 & 1.000 & 0.005 & 1.000 & -0.002 & 0.986 & 0.005 & 0.620\\
Sc.3 NR & Q-learning & -0.108 & 0.942 & 0.039 & 1.000 & 0.009 & 0.947 & 0.030 & 1.000\\
 & dWOLS & -0.108 & 0.940 & 0.039 & 1.000 & 0.009 & 0.669 & 0.030 & 1.000\\
 & BML-BP & -0.051 & 0.987 & 0.014 & 1.000 & -0.021 & 0.949 & 0.028 & 1.000\\
 & BML-OBART & -0.063 & 0.993 & 0.011 & 1.000 & 0.005 & 0.983 & 0.010 & 1.000\\
Sc.4 NNR & Q-learning & -0.098 & 0.948 & 0.037 & 0.773 & 0.008 & 0.947 & 0.029 & 0.759\\
 & dWOLS & -0.098 & 0.946 & 0.037 & 0.772 & 0.009 & 0.670 & 0.030 & 0.759\\
 & BML-BP & -0.038 & 0.990 & 0.013 & 0.717 & -0.022 & 0.948 & 0.028 & 0.759\\
 & BML-OBART & -0.055 & 0.994 & 0.010 & 0.826 & 0.004 & 0.981 & 0.010 & 0.847\\
Sc.5 NR & Q-learning & -0.003 & 0.968 & 0.031 & 0.663 & 0.023 & 0.950 & 0.038 & 1.000\\
 & dWOLS & -0.003 & 0.967 & 0.032 & 0.661 & 0.023 & 0.674 & 0.038 & 1.000\\
 & BML-BP & 0.024 & 0.983 & 0.022 & 0.621 & -0.016 & 0.951 & 0.035 & 1.000\\
 & BML-OBART & 0.022 & 0.999 & 0.008 & 0.710 & -0.009 & 0.972 & 0.019 & 1.000\\
Sc.6 RE & Q-learning & -0.015 & 0.964 & 0.028 & 1.000 & 0.004 & 0.941 & 0.032 & 0.999\\
 & dWOLS & -0.015 & 0.956 & 0.029 & 1.000 & 0.004 & 0.655 & 0.032 & 0.999\\
 & BML-BP & 0.045 & 0.971 & 0.027 & 1.000 & -0.011 & 0.945 & 0.029 & 0.999\\
 & BML-OBART & -0.009 & 0.997 & 0.008 & 1.000 & -0.002 & 0.971 & 0.014 & 1.000\\
Sc.7 RE & Q-learning & 0.020 & 0.956 & 0.032 & 0.880 & 0.018 & 0.943 & 0.034 & 1.000\\
 & dWOLS & 0.020 & 0.952 & 0.032 & 0.879 & 0.019 & 0.660 & 0.034 & 1.000\\
 & BML-BP & -0.035 & 0.973 & 0.018 & 0.884 & -0.015 & 0.946 & 0.031 & 1.000\\
 & BML-OBART & 0.023 & 0.998 & 0.008 & 0.990 & 0.004 & 0.979 & 0.015 & 1.000\\
Sc.8 NR & Q-learning & -0.092 & 0.951 & 0.036 & 0.990 & 0.003 & 0.945 & 0.025 & 1.000\\
 & dWOLS & -0.092 & 0.948 & 0.036 & 0.990 & 0.003 & 0.664 & 0.025 & 1.000\\
 & BML-BP & -0.145 & 0.892 & 0.041 & 0.992 & -0.048 & 0.929 & 0.025 & 1.000\\
 & BML-OBART & -0.054 & 0.992 & 0.010 & 1.000 & 0.004 & 0.990 & 0.008 & 1.000\\
Sc.9 NNR & Q-learning & -0.082 & 0.956 & 0.034 & 0.989 & 0.004 & 0.945 & 0.025 & 0.768\\
 & dWOLS & -0.082 & 0.953 & 0.034 & 0.989 & 0.004 & 0.663 & 0.025 & 0.769\\
 & BML-BP & -0.135 & 0.905 & 0.038 & 0.991 & -0.050 & 0.927 & 0.025 & 0.768\\
 & BML-OBART & -0.046 & 0.994 & 0.009 & 1.000 & 0.004 & 0.983 & 0.008 & 0.851\\
Sc.10 Squared & Q-learning & 0.035 & 0.416 & 0.618 & 0.423 & -0.055 & 0.453 & 0.480 & 0.532\\
 & dWOLS & 0.035 & 0.405 & 0.618 & 0.424 & -0.055 & 0.215 & 0.481 & 0.533\\
 & BML-BP & 0.100 & 0.248 & 0.608 & 0.425 & -0.038 & 0.225 & 0.458 & 0.532\\
 & BML-OBART & 0.024 & 0.903 & 0.203 & 0.937 & -0.044 & 0.817 & 0.201 & 0.914\\
Sc.11 Sin/Cos & Q-learning & 0.002 & 0.577 & 0.293 & 0.937 & 0.025 & 0.228 & 0.819 & 0.445\\
 & dWOLS & 0.003 & 0.545 & 0.293 & 0.937 & 0.025 & 0.108 & 0.820 & 0.445\\
 & BML-BP & 0.027 & 0.461 & 0.303 & 0.937 & 0.033 & 0.221 & 0.816 & 0.445\\
 & BML-OBART & -0.003 & 0.904 & 0.082 & 0.966 & 0.019 & 0.806 & 0.124 & 0.930\\
Sc.12 obs & Q-learning & 0.050 & 0.608 & 0.297 & 0.945 & 0.050 & 0.236 & 0.838 & 0.434\\
 & dWOLS & 0.013 & 0.634 & 0.295 & 0.941 & 0.021 & 0.107 & 0.838 & 0.428\\
 & BML-BP & 0.092 & 0.474 & 0.319 & 0.946 & 0.057 & 0.229 & 0.836 & 0.433\\
 & BML-OBART & 0.032 & 0.895 & 0.091 & 0.961 & 0.027 & 0.795 & 0.131 & 0.924\\
\bottomrule
\end{tabular}
}
\end{table}

\begin{table}[H]
\centering
\caption{The simulation results in terms of the `value' of the estimated DTR as compared to the observed and true optimal DTR based on the probability of getting the best outcome ($Y_2 = 3$) with $n_{tr} = 1000$. }
\label{tab4}
\resizebox{0.8\columnwidth}{!}{%
\begin{tabular}[t]{llcccccc}
\toprule
\multirow{2}{*}{Scenario} &  \multirow{2}{*}{Method} &\multicolumn{3}{c}{Stage 1} & \multicolumn{3}{c}{Stage 2} \\
&  & $P_{d^{*}}(Y=3)$ & $P_{\hat{d}^{*}}(Y=3)$ & $P(Y=3)$ & $P_{d^{*}}(Y=3)$ & $P_{\hat{d}^{*}}(Y=3)$ & $P(Y=3)$\\
\midrule
Sc.1 NR & Q-learning & 0.000 & 0.000 & 0.000 & 0.333 & 0.333 & 0.334\\
 & dWOLS & 0.000 & 0.000 & 0.000 & 0.333 & 0.333 & 0.334\\
 & BML-BP & 0.000 & 0.000 & 0.000 & 0.332 & 0.334 & 0.334\\
 & BML-OBART & 0.000 & 0.000 & 0.000 & 0.334 & 0.336 & 0.329\\
Sc.2 NNR & Q-learning & 0.000 & 0.000 & 0.000 & 0.337 & 0.333 & 0.334\\
 & dWOLS & 0.000 & 0.000 & 0.000 & 0.337 & 0.333 & 0.334\\
 & BML-BP & 0.000 & 0.000 & 0.000 & 0.336 & 0.334 & 0.334\\
 & BML-OBART & 0.000 & 0.000 & 0.000 & 0.337 & 0.332 & 0.332\\
Sc.3 NR & Q-learning & 1.000 & 1.000 & 1.000 & 0.527 & 0.528 & 0.401\\
 & dWOLS & 1.000 & 1.000 & 1.000 & 0.527 & 0.528 & 0.401\\
 & BML-BP & 1.000 & 1.000 & 1.000 & 0.526 & 0.528 & 0.401\\
 & BML-OBART & 1.000 & 1.000 & 1.000 & 0.525 & 0.528 & 0.401\\
Sc.4 NNR & Q-learning & 1.000 & 1.000 & 1.000 & 0.527 & 0.526 & 0.400\\
 & dWOLS & 1.000 & 1.000 & 1.000 & 0.527 & 0.526 & 0.400\\
 & BML-BP & 1.000 & 1.000 & 1.000 & 0.526 & 0.526 & 0.400\\
 & BML-OBART & 1.000 & 1.000 & 1.000 & 0.525 & 0.525 & 0.395\\
Sc.5 NR & Q-learning & 1.000 & 1.000 & 1.000 & 0.688 & 0.689 & 0.434\\
 & dWOLS & 1.000 & 1.000 & 1.000 & 0.688 & 0.689 & 0.434\\
 & BML-BP & 1.000 & 1.000 & 1.000 & 0.688 & 0.690 & 0.433\\
 & BML-OBART & 1.000 & 1.000 & 1.000 & 0.690 & 0.690 & 0.431\\
Sc.6 RE & Q-learning & 1.000 & 0.998 & 0.500 & 0.582 & 0.581 & 0.387\\
 & dWOLS & 1.000 & 0.998 & 0.500 & 0.582 & 0.581 & 0.387\\
 & BML-BP & 1.000 & 0.998 & 0.502 & 0.581 & 0.580 & 0.387\\
 & BML-OBART & 1.000 & 1.000 & 0.502 & 0.580 & 0.580 & 0.389\\
Sc.7 RE & Q-learning & 1.000 & 1.000 & 1.000 & 0.657 & 0.656 & 0.403\\
 & dWOLS & 1.000 & 1.000 & 1.000 & 0.657 & 0.656 & 0.403\\
 & BML-BP & 1.000 & 1.000 & 1.000 & 0.655 & 0.658 & 0.404\\
 & BML-OBART & 1.000 & 1.000 & 1.000 & 0.656 & 0.654 & 0.404\\
Sc.8 NR & Q-learning & 1.000 & 0.958 & 0.500 & 0.430 & 0.428 & 0.344\\
 & dWOLS & 1.000 & 0.957 & 0.500 & 0.430 & 0.428 & 0.344\\
 & BML-BP & 1.000 & 0.968 & 0.498 & 0.427 & 0.431 & 0.342\\
 & BML-OBART & 1.000 & 1.000 & 0.499 & 0.427 & 0.431 & 0.343\\
Sc.9 NNR & Q-learning & 1.000 & 0.956 & 0.500 & 0.433 & 0.430 & 0.344\\
 & dWOLS & 1.000 & 0.955 & 0.500 & 0.433 & 0.430 & 0.344\\
 & BML-BP & 1.000 & 0.967 & 0.498 & 0.430 & 0.431 & 0.343\\
 & BML-OBART & 1.000 & 1.000 & 0.500 & 0.430 & 0.429 & 0.343\\
Sc.10 Squared & Q-learning & 0.771 & 0.393 & 0.508 & 0.509 & 0.435 & 0.436\\
 & dWOLS & 0.771 & 0.395 & 0.508 & 0.509 & 0.435 & 0.436\\
 & BML-BP & 0.771 & 0.393 & 0.507 & 0.505 & 0.436 & 0.436\\
 & BML-OBART & 0.771 & 0.641 & 0.505 & 0.509 & 0.494 & 0.436\\
Sc.11 Sin/Cos & Q-learning & 0.998 & 0.952 & 0.535 & 0.527 & 0.439 & 0.434\\
 & dWOLS & 0.998 & 0.952 & 0.535 & 0.527 & 0.438 & 0.434\\
 & BML-BP & 0.998 & 0.953 & 0.535 & 0.525 & 0.438 & 0.433\\
 & BML-OBART & 0.998 & 0.983 & 0.534 & 0.527 & 0.522 & 0.431\\
Sc.12 obs & Q-learning & 0.998 & 0.955 & 0.472 & 0.514 & 0.422 & 0.419\\
 & dWOLS & 0.998 & 0.951 & 0.472 & 0.514 & 0.423 & 0.419\\
 & BML-BP & 0.998 & 0.957 & 0.475 & 0.511 & 0.424 & 0.418\\
 & BML-OBART & 0.998 & 0.988 & 0.474 & 0.513 & 0.505 & 0.419\\
\bottomrule
\end{tabular}
}
\end{table}

\newpage

\section{Application results}
\begin{figure}[H]
    \centering
     \renewcommand{\thesubfigure}{\Alph{subfigure}}
    \subfloat[]{%
        \includegraphics[width=0.5\linewidth]{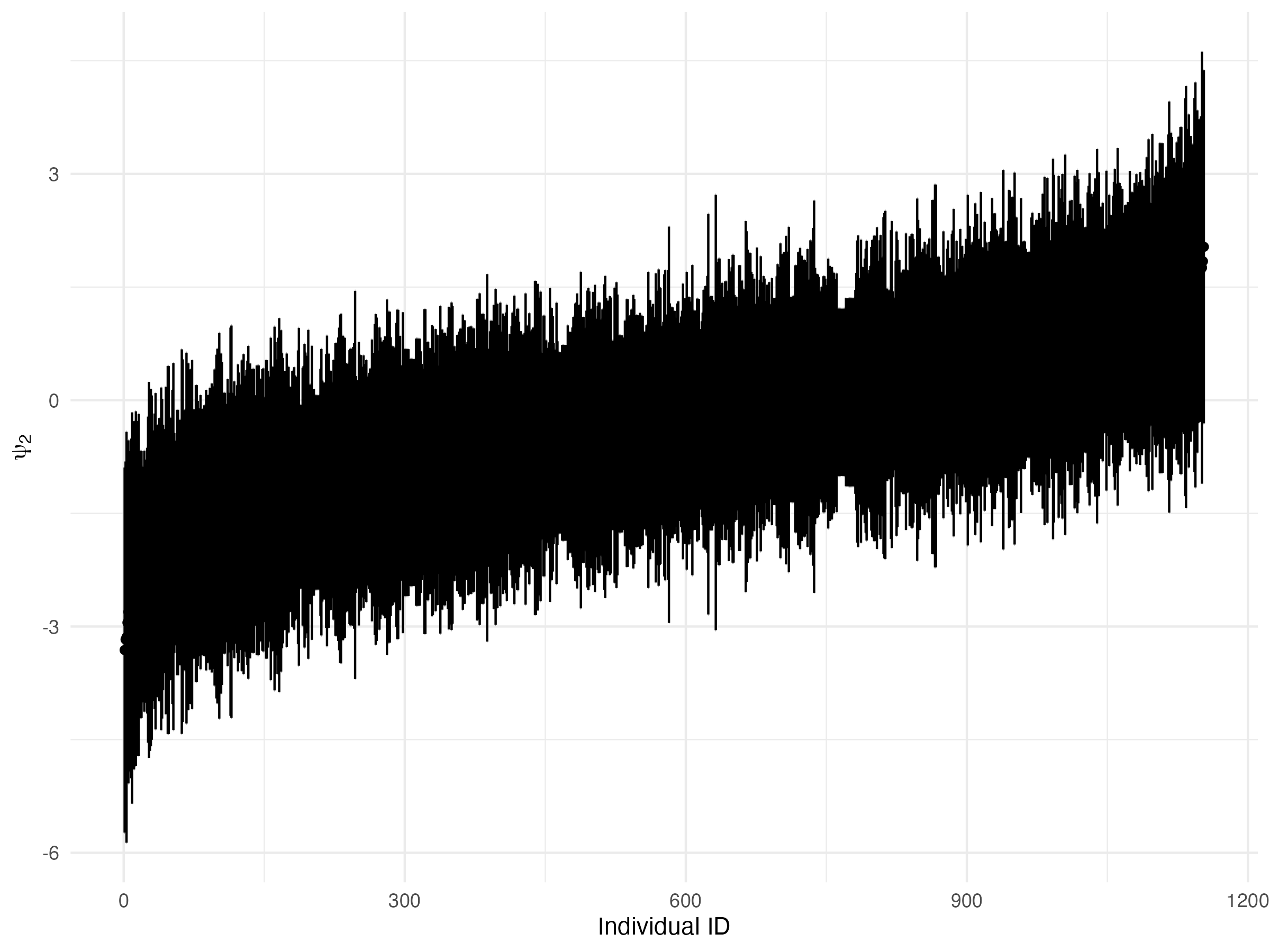}%
        }%
    \hfill%
    \subfloat[]{%
        \includegraphics[width=0.5\linewidth]{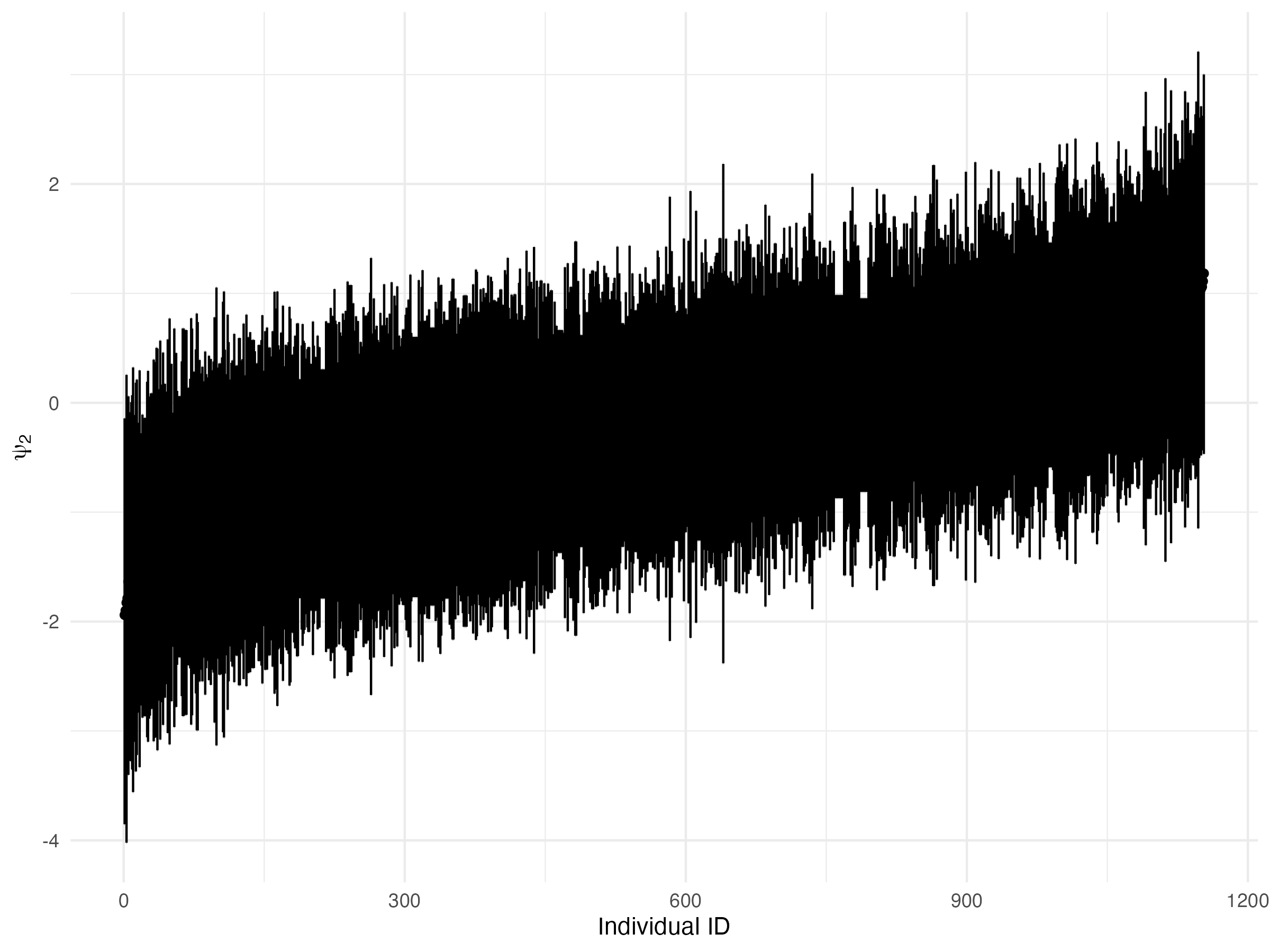}%
        }%
    \subfloat[]{%
        \includegraphics[width=0.5\linewidth]{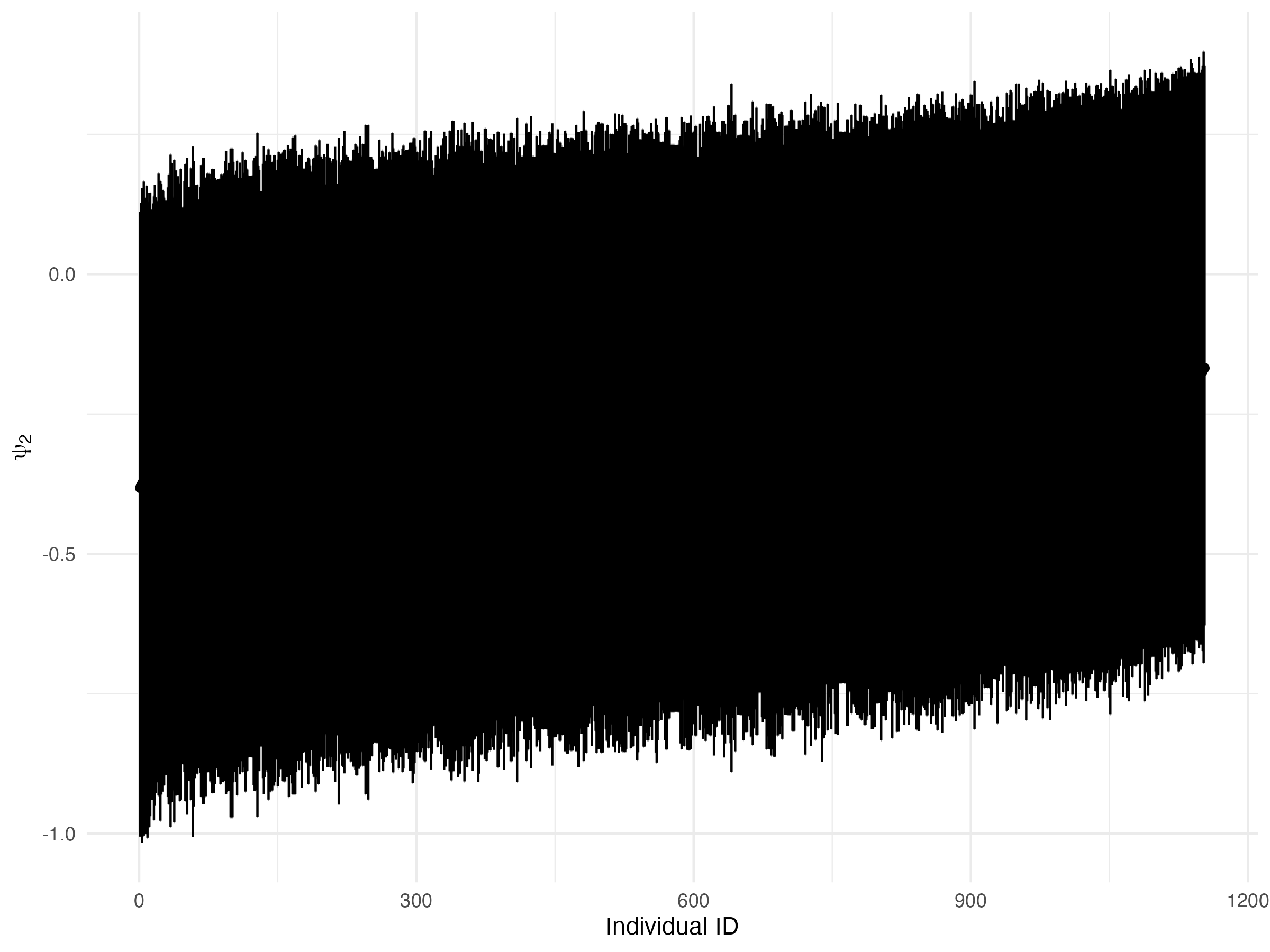}%
        }%
    \caption{Application results for estimating the distribution of $\psi_2$ using: A) Q-learning, B) the BML approach with ordered probit model, BML-BP, and C) the BML approach with OBART model, BML-OBART. }
    \label{appfig3}
\end{figure}

\begin{figure}[H]
    \centering
   \renewcommand{\thesubfigure}{\Alph{subfigure}}
    \subfloat[]{%
        \includegraphics[width=0.5\linewidth]{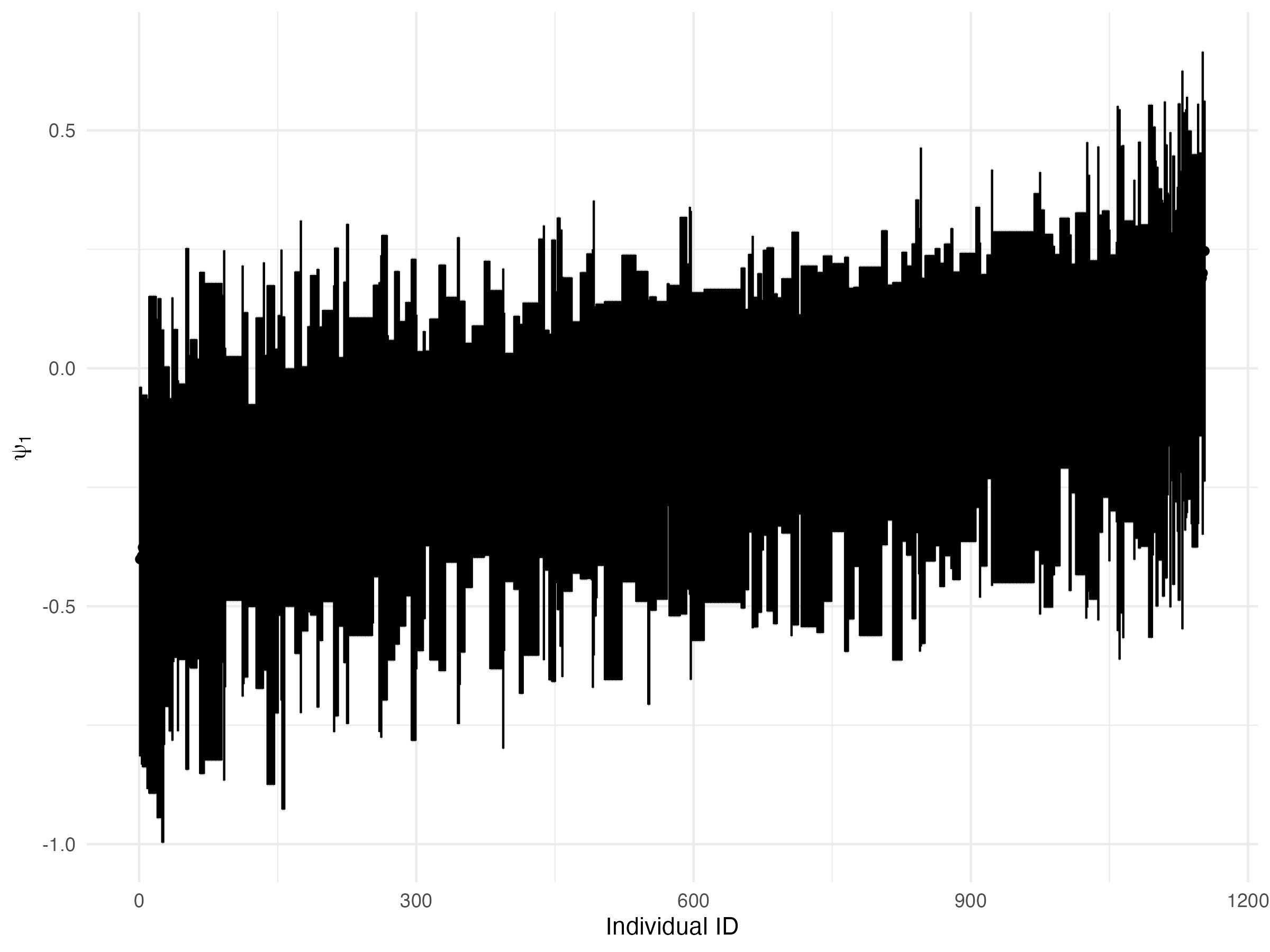}%
        }%
    \hfill%
    \subfloat[]{%
        \includegraphics[width=0.5\linewidth]{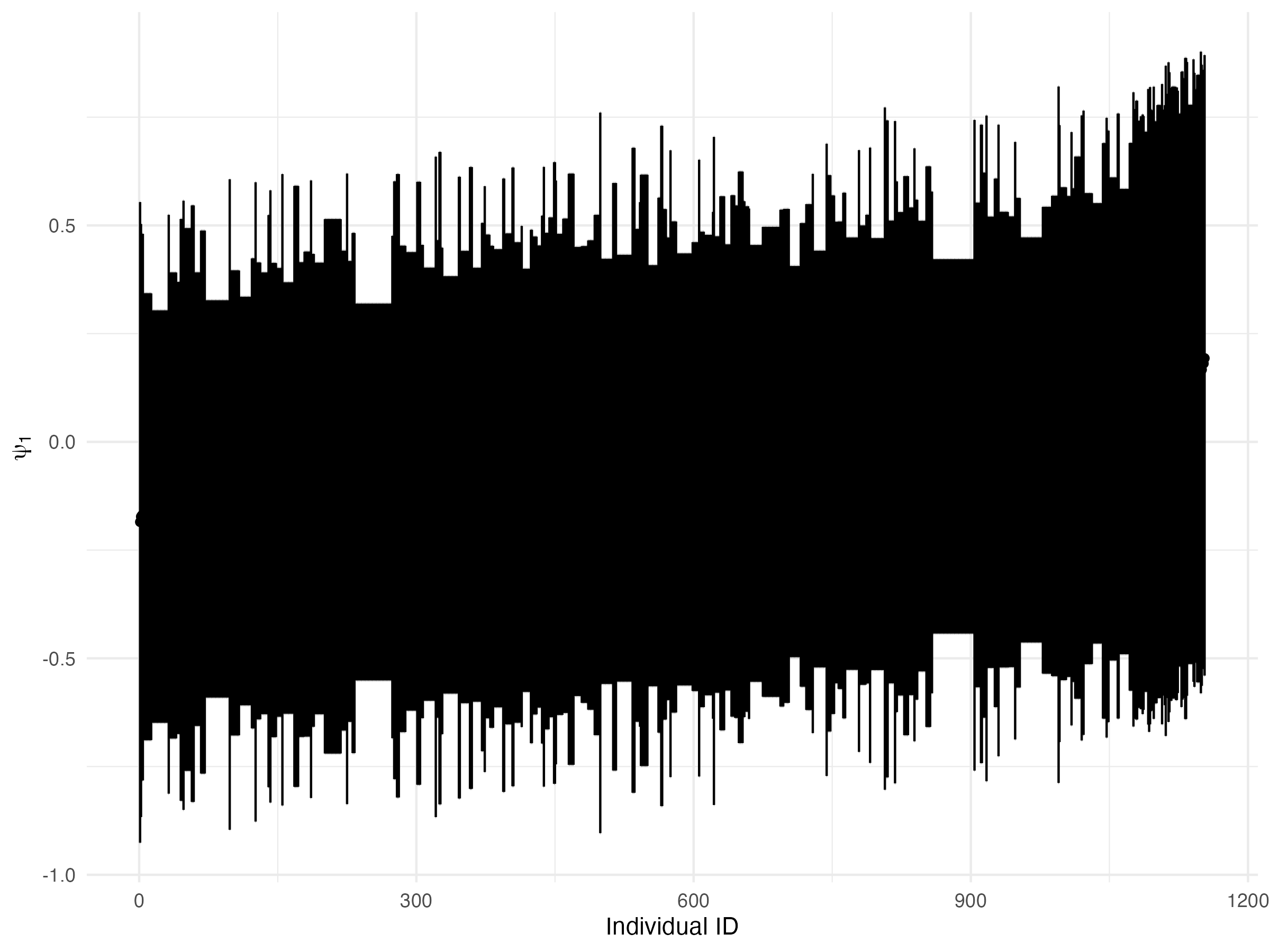}%
        }%
    \subfloat[]{%
        \includegraphics[width=0.5\linewidth]{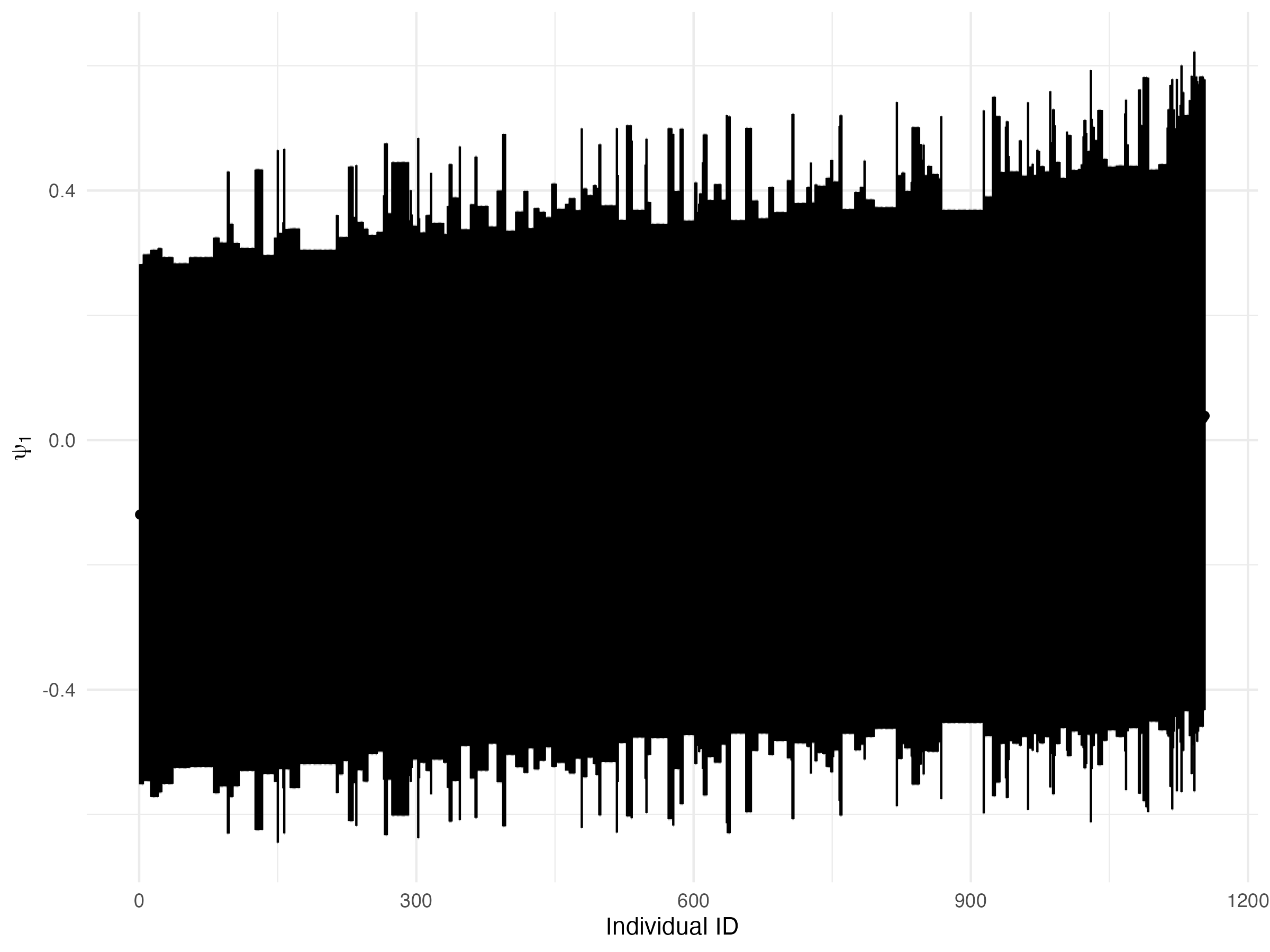}%
        }%
    \caption{Application results for estimating the distribution of $\psi_1$ using: A) Q-learning, B) the BML approach with ordered probit model, BML-BP, and C) the BML approach with OBART model, BML-OBART. }
    \label{appfig4}
\end{figure}